\documentclass[]{IEEEtran}
\usepackage[left=1.57cm,right=1.57cm,top=1.52cm,bottom=3cm]{geometry}
\IEEEoverridecommandlockouts

\usepackage{cite}
\usepackage{tikz}
\usetikzlibrary{arrows.meta}
\usepackage{pgfplots}
\usepackage{psfrag}
\usepackage{soul}
\usepackage{graphicx}
\usepackage{textcomp}
\usepackage{xcolor}
\usepackage{bbm}
\usepackage{subcaption}
\usepackage{amsmath,amssymb,amsfonts, acronym, amsthm}
\usepackage{hyperref}
\usepackage{xparse}
\usepackage{mathabx}

\usepackage{float}
\usepackage{amsmath, amssymb, bm, mathrsfs, yfonts}
\usepackage{algorithm}
\usepackage{algpseudocode}

\usepackage{color}
\usepackage[markup=underlined,ulem=normalem,xcolor=red]{changes}

\definechangesauthor[name=Alberto Toccafondi, color=orange]{AT}

\usepackage[singlespacing]{setspace} 

\definecolor{silver}{rgb}{0.95, 0.95, 0.95}

\IEEEoverridecommandlockouts

\usetikzlibrary{positioning} 
\definecolor{Eored}{rgb}{.647 ,.129 ,.149} 
\definecolor{Eogreen}{rgb}{0 ,0.53 ,0}

\newcommand{\REV}[1]{\textcolor{black}{{#1}}}

\begin{document}
\author{Andrea~Abrardo, \IEEEmembership{Senior~Member,~IEEE}, Giulio Bartoli \IEEEmembership{Member,~IEEE}, Alberto Toccafondi, \IEEEmembership{Senior~Member,~IEEE}
\thanks{The authors are with the University of Siena and CNIT. Email: abrardo@unisi.it, giulio.bartoli@unisi.it, alberto.toccafondi@unisi.it.}
}
\title{A Novel Comprehensive Multiport Network Model for Stacked Intelligent Metasurfaces (SIM) Characterization and Optimization}
\maketitle
\begin{abstract}
Reconfigurable Intelligent Surfaces (RIS) represent transformative technologies for next-generation wireless communications, offering advanced control over electromagnetic wave propagation.
While RIS have been extensively studied, Stacked Intelligent Metasurfaces (SIM), which extend the RIS concept to multi-layered systems, present significant modeling and optimization challenges.
This work addresses these challenges by introducing an optimization framework for SIM that, unlike previous approaches, is based on a comprehensive model without relying on specific assumptions, allowing for broader applicability of the results.
We first present a model based on multi-port network theory for characterizing a general electromagnetic collaborative object (ECO) and derive a framework for ECO optimization.
We then introduce the SIM as an ECO with a specific architecture, offering insights into SIM optimization for various configurations and discussing the complexities associated with each case.
\REV{Finally, we demonstrate that the comprehensive model considered in this work simplifies to the model traditionally used in the literature when the assumption of unilateral propagation between the levels of the SIM is made, and mutual coupling between the SIM elements is neglected.
To assess the applicability of these assumptions, a case study focused on the realization of a 2D DFT was undertaken.
In this context, we highlight that these assumptions introduce a significant mismatch between the SIM model and its behavior as described by the complete model, making these approximations inadequate for optimizing the SIM.
Conversely, we show that employing the complete model proposed in this paper can yield excellent performance.} 
\end{abstract}

\section{Introduction}

Reconfigurable Intelligent Surfaces (RIS) represent an innovative technology for next-generation wireless networks, particularly in the context of mmWave frequencies \cite{DiRenzo:19b,WuZhaZheYouZha:20,DiRenzo2020_JSAC}.
Most optimization works model RIS as planar arrays of reflective elements whose impedances can be adjusted to create controllable phase-shifts, shaping the reflected wavefront.
However, these models often lack electromagnetic consistency, not fully accounting for factors critical to realistic RIS operation \cite{RenzoDT22, AbeywickramaZWY20}.
Recent advancements highlight the necessity of accurate reradiation models, combining surface-level optimization with precise design of RIS elements \cite{RafiqueHZNMRDY23}.
Multiport network theory has emerged as an effective method for ensuring the accuracy of models while enabling easy system-level optimization \cite{IvrlacN10,DR1,DR2,ABR_MUTUAL,RenzoZDAYRT20,10213362}.
New approaches based on S and Z parameters reveal the limitations of classical RIS models that treat them as ideal scatterers, often neglecting important aspects such as electromagnetic mutual coupling, the presence of unwanted reflections, and the correlation between reflection coefficient phase and amplitude \cite{Abrardo24RisOpt}.
Incorporating these factors leads to more robust end-to-end models capable of optimizing all scattering components and minimizing unwanted interferences.

Initial research on RIS primarily focused on single-connected, reflective RIS models, characterized by diagonal phase-shift matrices.
However, the limitations of these simplified models, particularly in terms of flexibility and scalability, have motivated the development of more advanced metasurface-based systems.
Among these, particular emphasis is placed on the so-called beyond-diagonal structures, in which different ports of the RIS are interconnected through programmable lines, creating more complex and flexible structures \cite{10155675}.
In this context, in addition to traditional purely reflective RIS, hybrid transmissive and reflective structures have also been considered, which are referred to as simultaneously transmitting and reflecting RIS (STAR-RIS) \cite{Li_2023_ref1}.
Just as with classical RIS, in these more complex architectural scenarios, the traditional approach to managing the complexity of optimization typically involves relying on certain assumptions, which can limit the generality of the results.

Recently, a novel technology relying on stacked intelligent metasurfaces (SIM) has emerged by cascading multiple transmitting RIS (T-RIS) \cite{DiffractiveDNN}, which is capable of implementing signal processing in the EM wave regime.
This represents a significant advancement, providing improved control over wave propagation and greatly increasing the degrees of freedom \cite{DBLP:journals/jsac/AnXNAHYH23}.
In a SIM, each intelligent metasurface acts like a layer in a Deep Neural Network (DNN), while each programmable meta-atom functions similarly to a neuron, possessing adjustable phase and amplitude responses that can be tailored to meet various task needs and adapt to changing environments.
Consequently, SIM benefits from the strong representation capabilities of Artificial Neural Networks (ANNs), the exceptional speed of electromagnetic (EM) computing, and the energy-efficient tuning properties of metasurfaces.

Although the literature on SIM is still limited, it is rapidly expanding due to the significant interest in this topic. SIMs have been shown to effectively perform beamforming in the electromagnetic domain and to implement holographic multiple-input multiple-output communications without requiring excessive radio-frequency (RF) chains  \cite{DBLP:journals/ojcs/HassanARDY24,DBLP:journals/jsac/AnXNAHYH23}. Moreover, SIM can be used to enhance the performance of multi-user beamforming \cite{DBLP:journals/wc/AnYXLNRDH24,DBLP:journals/vtm/BasarALWJYDS24,DBLP:conf/icc/AnRDY23}.
In \cite{DBLP:journals/wcl/PapazafeiropoulosKCKV24} the achievable rate of a large SIM-aided system with statistical CSI is derived and an optimization procedure based on AO is proposed. In \cite{DBLP:conf/icc/LiuANA024} a deep reinforcement learning approach is proposed to overcome limitations of traditional AO approaches.
The use of SIM in cell-free networks is explored in \cite{CellFreeDLSIM} for the downlink and in \cite{DBLP:journals/tcom/LiEXAYH24} for the uplink, where the multi-user beamforming is designed for a system where each AP has its own SIM. Moreover, the work in \cite{DBLP:journals/wcl/LinAGDY24} considers a LEO satellite equipped with a SIM.
Some works focus on near field communications, such as \cite{DBLP:journals/wcl/PapazafeiropoulosKCKV24a} where users are equipped with multiple antennas, and \cite{DBLP:conf/vtc/JiaALGRDY24}, where the diffraction behavior of SIM meta-atoms is taken into account. Furthermore, SIMs can be used to improve sensing performance, as for example the estimation of direction of arrival can be enhanced by the use of the SIM, as analyzed in \cite{DBLP:conf/icc/AnY0RDPH24} and \cite{DBLP:journals/jsac/AnYGRDPH24}, and the use of SIM for integrated sensing and communications (ISAC) problems has been studied in \cite{DBLP:journals/wcl/NiuAPGCD24,SIM_ISAC,AntiJammingSIM}.

Despite the promise of SIM, accurate and tractable modeling remains a significant challenge. Existing research employs a simple model in which the SIM is characterized as a cascade, consisting of the propagation through the channels that separate the layers, along with the phase shifts introduced during the transition through each layer. An important effort to provide a more accurate model, highlighting the intrinsic approximations in the simplified model previously used for SIM optimization, is presented in \cite{Nerini_Clerckx_SIM}. 

\REV{In \cite{Nerini_Clerckx_SIM} an S-parameters representation of the SIM is provided in which the SIM is modeled as a cascade of $L$ blocks, each consisting of the cascade of a wireless channel and an RIS. This reflects the choice of incorporating both the wireless channel and the load network of the SIM-layer into each block’s S-matrix representation. Consequently, the controllable parameters of each SIM-layer are embedded within the S-matrix representation of that block. However, the scattering matrix representation complicates handling cascaded electromagnetic problems due to the nature of the electrical quantities defined at each port. This difficulty impacts the total channel matrix representation and it is hard to understand the role of the reconfigurable scattering matrices. To achieve a more tractable representation, the authors simplify the channel model by assuming ideal T-RISs without mutual coupling and employing a unilateral approximation for propagation through the channels separating the layers of the constituent T-RISs in the SIM. This approximation, facilitate the mathematical treatment but may limit the model’s applicability to practical, non-reciprocal propagation environments. Under these assumptions, the model in \cite{Nerini_Clerckx_SIM}  effectively reduces to the commonly adopted cascade structure with phase shifts at each layer, which is also employed in optimization frameworks. Essentially, \cite{Nerini_Clerckx_SIM} highlights the limitations of this model, but it still considers this model for SIM optimization.}

\REV{It remains an open question to assess the applicability of such assumptions and, if they do not hold, to verify whether a SIM can still achieve the promising performance indicated by initial studies.} \REV{To this aim, in our proposed approach, we consider a Z-parameters representation of the SIM. This representation yields an equivalent complete input-output model of the system as the S-parameters representation presented in \cite{Nerini_Clerckx_SIM}, allowing us to model the problem as a global interaction between the SIM, the transmitter, and the receiver. The Z-parameters model facilitates the representation of the transfer function through band matrices, thereby enabling an iterative approach for evaluating the gradient, which is essential for optimizing the SIM, even for the general SIM model without approximations.}

\subsection{Contributions}
This work aims to provide a comprehensive analysis of SIM systems. The main contributions are:
\begin{enumerate}
    \item \textbf{General ECO Model:} A thorough multiport network model of a general electromagnetic collaborative object (ECO) is introduced, generalizing previous models for diagonal RIS, non-diagonal RIS, STAR-RIS, and SIM. Then, an optimization procedure based on gradient descent is developed without relying on specific assumptions or approximations.
    \item \textbf{Complete SIM Model:} The general ECO model is specialized for the SIM case. Therefore, the gradient-descent-based optimization approach is tailored to the SIM case. We derived an iterative algorithm for gradient calculation that appropriately leverages the layered architecture of the SIM and allows for a significant complexity reduction compared to the general ECO case. This can be summarized as follows: instead of being constrained by the total number of elements in the SIM, it is only constrained by the number of elements in each layer.
    \item \textbf{Simplified SIM Models and Backpropagation Algorithm:}  Several simplifications of the general SIM model are analyzed, leading to the case of unilateral approximation and ideal diagonal T-RIS. In this latter case, we demonstrate how the general Z-parameter model developed here turns out to be the same as those previously considered in the literature. Additionally, a backpropagation-based algorithm is provided, allowing for complexity reduction by exploiting the characteristics of the simplified model.
    \item \textbf{Performance evaluation considering different SIM models:} \REV{We demonstrate how the proposed framework enables the optimization of the SIM with diagonal constituent T-RISs for implementing a 2D DFT. In this context, we highlight that the assumptions typically made in previous works introduce a significant mismatch between the SIM model and its behavior as described by the complete model, making these approximations inapplicable for optimizing the SIM. Conversely, we show that employing the complete model proposed in this paper can yield excellent performance.} Notably, the results indicate that the model without approximations—being more complete and realistic—can yield better performance than the simplified model without mismatch. This outcome appears to depend on the presence of some coupling among the elements of the T-RIS and a feedback effect between the layers, which is absent when the unilateral approximation is applied. This allows for greater design flexibility, even when T-RISs are implemented with simple diagonal architectures.    
\end{enumerate}
\subsection{Paper Outline and Notation}
The remainder of this paper is structured as follows. In Section \ref{sec:General_model}, we present the general multiport model of an ECO. In Section \ref{PROC_EM}, we propose an optimization framework based on a gradient descent approach for an ECO. In Section \ref{SIM_model}, we focus on a SIM, deriving the model and the optimization framework for this special case. Finally, in Section V, we provide the results and comparisons.

{\sl Notation}: Matrices are denoted by bold uppercase letters (i.e., $\mathbf{X}$), vectors are represented by bold lowercase letters (i.e., $\mathbf{x}$), and scalars are denoted by normal font (i.e., $x$). $(\cdot)^{\mathrm{T}}$, $(\cdot)^{\mathrm{H}}$, $(\cdot)^{-1}$ and tr() stand for the transpose, Hermitian transpose, inverse and trace of the matrices. The symbol $\text{diag}\left(\mathbf{x}\right)$ is the diagonal matrix obtained from the element of vector $\mathbf{x}$. Finally, $\mathbf{I}_n$ indicates the identity matrix of dimension $n$.

\section{General Multi-port model}\label{sec:General_model}
\begin{figure}[h!]
    \centering    \includegraphics[width=\columnwidth]{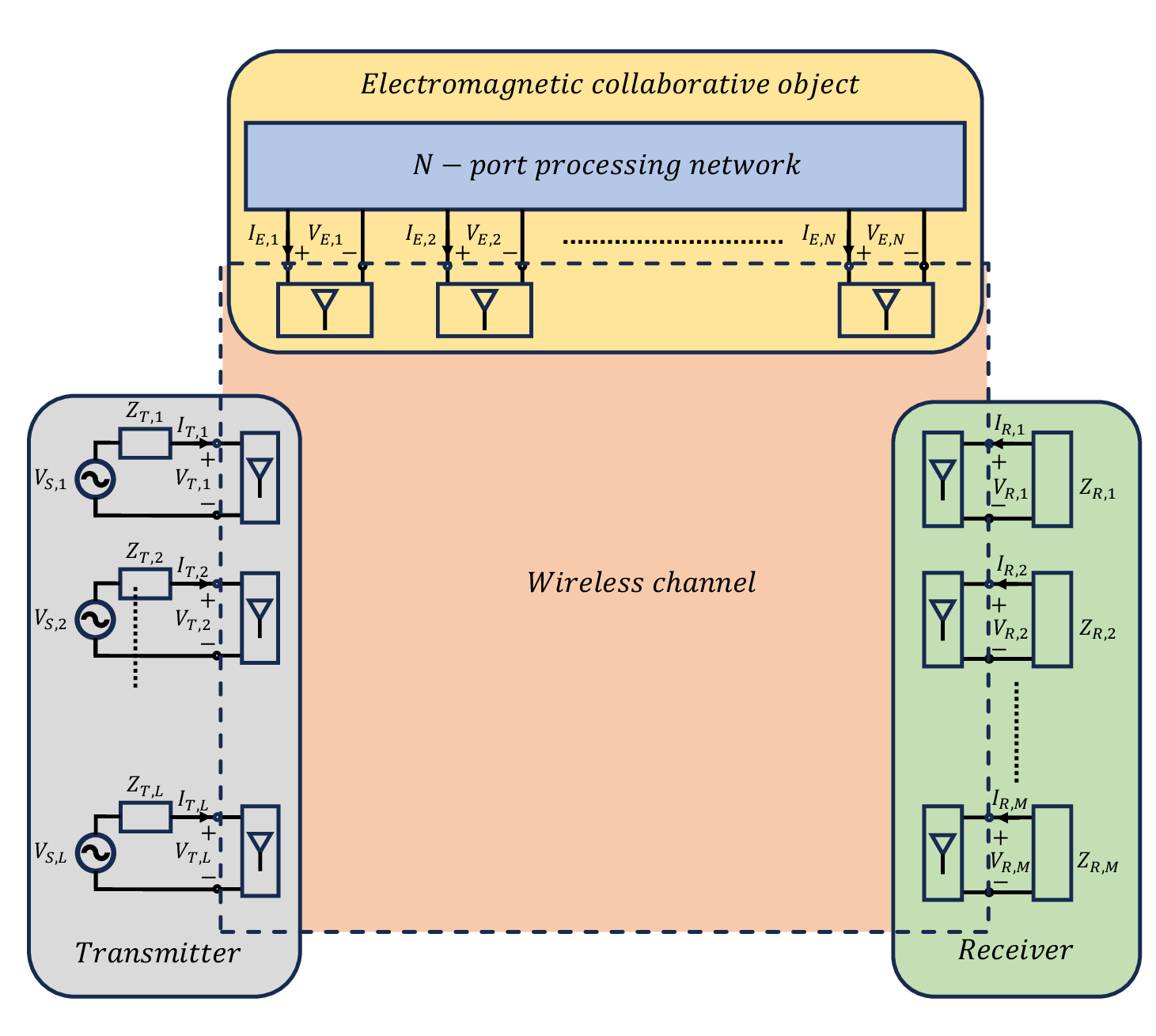}
    \caption{Network model.}  \label{Fig_1_network} \end{figure}

Let's consider the multiport system model shown in Fig.
 \ref{Fig_1_network}, which is a general framework for characterizing a transmitter with $L$ ports (e.g., a multi-antenna transmitter), a receiver with $M$ ports (e.g., a multi-antenna receiver), along with $N$ ports corresponding to $N$ elements of an object that receives, processes, and retransmits electromagnetic waves to and from the wireless channel. This object can generically represent a RIS, whether reflective or transmitting or operating in both modes, such as a Simultaneously Transmitting And Reflecting RIS (STAR-RIS), or a Stacket Intelligent Metasurfaces (SIM). To remain general, let's call this object an ElectroMagnetic Collaborative Object (ECO). The Z-parameters representation of the multiport network relates the voltages $V$ and the currents $I$ at the ports as follows:
\begin{equation}
    \left[\begin{array}{c}\mathbf{V}_T \\\mathbf{V}_E \\\mathbf{V}_R\end{array}\right]= 
    \left[\begin{array}{ccc}\mathbf{Z}_{TT} & \mathbf{Z}_{TE}& \mathbf{Z}_{TR}\\
    \mathbf{Z}_{ET} & \mathbf{Z}_{EE}& \mathbf{Z}_{ER}\\
    \mathbf{Z}_{RT}& \mathbf{Z}_{RE}& \mathbf{Z}_{RR}
    \end{array}\right] 
    \left[\begin{array}{c}\mathbf{I}_T \\\mathbf{I}_E \\\mathbf{I}_R\end{array}\right] \nonumber,
    \label{eq:absub}
  \end{equation}
where $\mathbf{I}_x$ and $\mathbf{V}_x$ for $x \in \{T,E,R\}$ denote the currents and voltages at the ports of the transmitter ($T$), ECO ($E$), and receiver ($R$).  
Moreover, the voltages and the currents at the ECO ports are related as $\mathbf{V}_E = -\mathbf{Z}_E \mathbf{I}_E$, where $\mathbf{Z}_E$ is the impedence matrix of the network to which the ECO ports are connected.

The presented model is a generic multi-port network model that has been studied in the literature, primarily in the context of RIS, for which the transfer function $\mathbf{V}_R = \mathbf{{H}}_{Z}\mathbf{V}_T$ is known. In particular, under the conditions shown in \cite{Abrardo24RisOpt}, we have:
\begin{equation}\label{eq:TFZ}
    \mathbf{{H}}_{Z} = \frac{1}{4Z_0}\left[\mathbf{Z}_{RT}-\mathbf{Z}_{RE} (\mathbf{Z}_{EE}+\mathbf{Z}_{E})^{-1}\mathbf{Z}_{ET}\right].
  \end{equation}
where $Z_0$ is the characteristic impedance to which both the transmitter and the receiver are matched. This model therefore generally represents the relationship between the output and input signals in a system containing an ECO. \REV{It is important to note that the $\mathbf{Z}-$matrix representation of an electromagnetic system provides a complete network representation, accounting for both self-impedances (diagonal elements, $Z_{nn}$) at ports and mutual impedances (off-diagonal elements, $Z_{mn}$) between ports. When analyzing the coupling between two antennas with accessible ports, these mutual impedances are determined based on the total field contributions at each antenna port. 
Consequently, the Z-matrix representation inherently includes near-field coupling effects, as it models the full electromagnetic interaction between antennas, whether they are in the radiative far-field zone or strongly coupled in the near-field region.}

Note that if the ECO is an RIS and the network connecting the RIS elements ensures each element has its own termination impedance without interconnections between them, the RIS operates as a classical diagonal RIS. On the other hand, if the network also includes connections between different RIS elements, we have a non-diagonal RIS also called beyond-diagonal RIS (BD-RIS) \cite{10155675}. Furthermore, if the load network allows some RIS elements to let the signal pass through to other RIS elements, the same model can describe a RIS operating in transmissive mode, where some elements receive the signal and others transmit it \cite{Li_2023_ref1}. Finally, if we divide the ECO ports into electromagnetically isolated groups comprising some receiving ports and some transmitting ports, we can also describe a SIM. In this case, each group is a T-RIS representing a layer of the SIM. The model is thus general, and to use it in optimizing the operation of the ECO it is necessary to adequately handle the nonlinear part of the transfer function, which includes the matrix inversion and depends on the controllable or tunable parameters of the ECO, i.e., the load network $\mathbf{Z}_E$.
\section{Processing in the Electromagnetic Domain}\label{PROC_EM}
Let us assume that the load network depends on a certain number $P$ of controllable parameters and let us denote $\boldsymbol{\eta} \in \mathbb{C}^{P\times1}$ the vector of controllable parameters, with the resulting $\mathbf{Z}_E$ expressed as $\mathbf{Z}_E(\boldsymbol{\eta})$. To remain general, let then $\mathbf{A} \in \mathbb{C}^{M \times N}$ be a matrix which may include the receiver-ECO impedance matrix $\mathbf{Z}_{RE}$, as well as a linear filter used to extract an estimate of the transmitted signal. It can also be utilized to implement generic processing of the signal received by the ECO by, for example, using $M$ probes to detect the received signal. Additionally, let $\mathbf{b} \in \mathbb{C}^{N \times 1}$ be the vector received at the ECO. 
Moreover, 
define $\mathbf{T}(\boldsymbol{\eta}) = (\mathbf{Z}_{EE} + \mathbf{Z}_{E}(\boldsymbol{\eta}))^{-1}$ and consider
\begin{equation}
\label{eqdef2}
\begin{aligned}
\mathbf{h}_T(\boldsymbol{\eta}) = \mathbf{A} \mathbf{T}(\boldsymbol{\eta}) \mathbf{b}.
\end{aligned}
\end{equation} 

Hence, the expression in \eqref{eqdef2} represents a generic transfer function that includes the effect of the ECO and can be used to optimize it in various application scenarios. In the following, we present a generic ECO optimization problem that can be adapted to various contexts of electromagnetic processing. In particular, we consider the problem of designing the ECO so that, given a set of $I$ inputs $\mathbf{b}_i$, $i = 1,2,\ldots,I$, it provides an output that \emph{closely approximates} the outputs $\mathbf{x}_i \in \mathbb{C}^{M \times 1}$, $i = 1,2,\ldots,I$. To this aim, let first introduce $\mathbf{{h}}^{(i)}_{T} = \mathbf{A} \mathbf{T}(\boldsymbol{\eta}) \mathbf{b}_i$ and denote by $\epsilon_i(\boldsymbol{\eta})$ the $i$-th squared error:
\begin{equation}
\label{eq:mMSEMat0_1}
\begin{aligned}
\epsilon_i(\boldsymbol{\eta}) = \left(\mathbf{{h}}^{(i)}_{T}(\boldsymbol{\eta})-{\mathbf{x}_i}\right)^H\left(\mathbf{{h}}^{(i)}_{T}(\boldsymbol{\eta})-{\mathbf{x}_i}\right).
\end{aligned}
\end{equation}

Elaborating from \eqref{eq:mMSEMat0_1}, we get:
\begin{equation}
\label{eq:mMSEMat0_2}
\begin{aligned}
\epsilon_i(\boldsymbol{\eta}) =\left(\mathbf{{h}}^{(i)}_{T}(\boldsymbol{\eta})\right)^H\mathbf{{h}}^{(i)}_{T}(\boldsymbol{\eta}) -2\Re\left(\mathbf{x}_i^H\mathbf{h}^{(i)}_{T}(\boldsymbol{\eta})\right)+\mathbf{x}_i^H\mathbf{x}_i.
\end{aligned}
\end{equation}

We then consider the following problem:
\begin{align}
\label{P:mineps}
\min \limits_{\boldsymbol{\eta}} \sum\limits_i\epsilon_i(\boldsymbol{\eta}).
\end{align}

To find an efficient strategy for solving the problem \eqref{P:mineps}, it is necessary to calculate the gradient $\nabla_{\boldsymbol{\eta}}\epsilon_i(\boldsymbol{\eta})$ which depends on the evaluation of terms of the form $\nabla_{\boldsymbol{\eta}} \mathbf{x}_i^H\mathbf{h}^{(i)}_T(\boldsymbol{\eta})$ and $\nabla_{\boldsymbol{\eta}}\left(\mathbf{h}^{(i)}_T(\boldsymbol{\eta})\right)^H\mathbf{h}^{(i)}_T(\boldsymbol{\eta})$.

To elaborate, let us define by $\mathbf{G}_p(\boldsymbol{\eta}) = \frac{\partial \mathbf{Z}_{E}(\boldsymbol{\eta})}{\partial \eta_p} \in \mathbb{C}^{N \times N}$ the tangent matrix of $\mathbf{Z}_{E}(\boldsymbol{\eta})$ with respect to $\eta_p$, i.e., the matrix obtained by evaluating the element-wise partial derivative of $\mathbf{Z}_{E}(\boldsymbol{\eta})$ with respect to $\eta_p$. Then, introduce:
\begin{equation}
\label{eq:dp_fp}
\begin{aligned}
d^{(i)}_p(\boldsymbol{\eta}) & = \frac{\partial ~\mathbf{x}_i^H\mathbf{h}^{(i)}_T(\boldsymbol{\eta})}{\partial \eta_p},\quad
f^{(i)}_p(\boldsymbol{\eta}) & = \frac{\partial ~\left(\mathbf{h}^{(i)}_T(\boldsymbol{\eta})\right)^H \mathbf{h}^{(i)}_T(\boldsymbol{\eta})}{\partial \eta_p}
\end{aligned}
\end{equation}
and the vectors $\mathbf{d}^{(i)}(\boldsymbol{\eta}) \in \mathbb{C}^{1 \times P}$ and $\mathbf{f}^{(i)}(\boldsymbol{\eta}) \in \mathbb{C}^{1 \times P}$ that contain in $p$-th position $d^{(i)}_p(\boldsymbol{\eta})$ and $f^{(i)}_p(\boldsymbol{\eta})$, respectively. From the Neumann series expansion of the inverse of matrices is possible to derive from \eqref{eqdef2}:
\begin{equation}
\label{eq:Grad1}
\begin{aligned}
d^{(i)}_p(\boldsymbol{\eta}) & =  -\mathbf{x}_i^H\mathbf{A}\mathbf{T}(\boldsymbol{\eta})\mathbf{G}_p(\boldsymbol{\eta})\mathbf{T}(\boldsymbol{\eta})\mathbf{b}_i\\
f^{(i)}_p(\boldsymbol{\eta}) & =  -2\Re\left\{\left(\mathbf{h}^{(i)}_T(\boldsymbol{\eta})\right)^H\mathbf{A}\mathbf{T}(\boldsymbol{\eta})\mathbf{G}_p(\boldsymbol{\eta})\mathbf{T}(\boldsymbol{\eta})\mathbf{b}_i\right\}.
\end{aligned}
\end{equation}

Due to the nonlinearity of the function $\mathbf{T}(\boldsymbol{\eta})$, the problem \eqref{P:mineps} is non-convex; therefore, it is necessary to develop a suboptimal strategy to find a local minimum. To this end, leveraging \eqref{eq:Grad1}, the gradient descent algorithm can be employed. 
To elaborate, $\boldsymbol{\eta}$ can be adjusted iteratively according to:
\begin{equation}
\label{eq:mMSEMat0_3}
\begin{aligned}
\boldsymbol{\eta}^{(q+1)} =  \boldsymbol{\eta}^{(q)} - \alpha \sum\limits_i \left[\mathbf{f}^{(i)}\left(\boldsymbol{\eta}^{(q)}\right) -2\Re\left( \mathbf{d}^{(i)}\left(\boldsymbol{\eta}^{(q)}\right)\right)\right],
\end{aligned}
\end{equation}
where $\alpha$ is the learning rate. The problem just described can be viewed as an example of ECO optimization using a supervised training set $(\mathbf{b}_i, \mathbf{x}_i)$, and can therefore be applied to classical scenarios of supervised learning. On the other hand, it is easy to see that the proposed approach can also be used to implement a known linear transfer function $\boldsymbol{\Theta} \in \mathbb{C}^{M \times N}$, such as that of MIMO beamforming or the calculation of the 2D DFT. This can be achieved by designing the ECO according to the following minimum square error criterion:
\begin{equation}
\label{eq:Transf1}
\begin{aligned}
\min \limits_{\boldsymbol{\eta}} \text{tr}\left[\left(\mathbf{A} \mathbf{T}(\boldsymbol{\eta})-{\boldsymbol{\Theta}}\right)\left(\mathbf{A} \mathbf{T}(\boldsymbol{\eta})-{\boldsymbol{\Theta}}\right)^H\right].
\end{aligned}
\end{equation}

If we denote the $i$-th column of $\mathbf{A} \mathbf{T}(\boldsymbol{\eta})$ as $\mathbf{y}_i(\boldsymbol{\eta})$ and the $i$-th column of $\boldsymbol{\Theta}$ as $\mathbf{x}_i$, with $i = 1, \ldots, N$, the criterion in \eqref{eq:Transf1} can be rewritten as:
\begin{equation}
\label{eq:Transf2}
\begin{aligned}
\min \limits_{\boldsymbol{\eta}} \sum\limits_{i } \left(\mathbf{y}_i(\boldsymbol{\eta})-\mathbf{x}_i\right)^H\left(\mathbf{y}_i(\boldsymbol{\eta})-\mathbf{x}_i\right).
\end{aligned}
\end{equation}

Since $\mathbf{y}_i(\boldsymbol{\eta}) = \mathbf{A} \mathbf{T}(\boldsymbol{\eta})\mathbf{e}_i$, where $\mathbf{e}_i \in \mathbb{C}^{N \times 1}$ is the vector of all zeros except in the $i$-th position, where it is one, problem \eqref{eq:Transf1} can be seen as a particular case of \eqref{P:mineps} when $I = N$ and $\mathbf{b}_i = \mathbf{e}_i$.
\subsection{Computational complexity}
Now we consider the computational complexity of the ECO optimization problem as the complexity due to each single iteration of the gradient descent algorithm. This same quantity will then be taken into account in subsequent cases when specific SIM architectures are considered.

To begin with, it is necessary to define the complexity of calculating $\mathbf{G}_p(\boldsymbol{\eta})$ for a generic $p$. This complexity strongly depends on the type of network considered for connecting the ports of the ECO, specifically on whether or not there exists an easily derivable analytical formulation for $\mathbf{Z}_E(\boldsymbol{\eta})$. For example, in the case where the ECO is a classical diagonal RIS, the matrix $\mathbf{Z}_E(\boldsymbol{\eta})$ is diagonal, and each element depends on a single variable parameter, such as the phase of the reflection coefficient at the port. In this scenario, the calculation is straightforward. In the case of BD-RIS, the calculation can be more complicated; however, if a closed-form and differentiable expression of $\mathbf{Z}_E(\boldsymbol{\eta})$ exists, the complexity of calculating $\mathbf{G}_p(\boldsymbol{\eta})$ can be neglected compared to the other calculations necessary for computing the terms $d^{(i)}_p(\boldsymbol{\eta})$ and $f^{(i)}_p(\boldsymbol{\eta})$ reported in \eqref{eq:Grad1}. A more accurate characterization of the calculation of $\mathbf{G}_p(\boldsymbol{\eta})$ will be provided later for a specific case of SIM characterized by diagonal RIS.

To elaborate, the evaluation of $d^{(i)}_p(\boldsymbol{\eta})$ and $f^{(i)}_p(\boldsymbol{\eta})$ can be accomplished following algorithm \ref{Algo1}.
\begin{algorithm}
    \caption{$ECO$: Evaluation of 
$d^{(i)}_p(\boldsymbol{\eta})$ and $f^{(i)}_p(\boldsymbol{\eta})$}\label{Algo1}
    \small
    \begin{algorithmic}[1]
        \State \textbf{Input:} $\boldsymbol{\eta}, \mathbf{A}, \mathbf{b}_i, \mathbf{x}_i, \mathbf{Z}_{EE},  \mathbf{Z}_{E}(\boldsymbol{\eta}), \mathbf{G}_p(\boldsymbol{\eta})$
        
        \State \text{Evaluate $\mathbf{T}(\boldsymbol{\eta}) = \left(\mathbf{Z}_{EE}+\mathbf{Z}_{E}(\boldsymbol{\eta})\right)^{-1}$ $\mathcal{O}\left(N^3\right)$}
        \State \text{Evaluate $\mathbf{J}_0 = \mathbf{A}\mathbf{T}(\boldsymbol{\eta})$ $\mathcal{O}\left(MN^2\right)$}
        
        \For{p = 1 \textbf{to} P}
            \State \text{Evaluate $\mathbf{J}_{1,p} = \mathbf{J}_0\mathbf{G}_p(\boldsymbol{\eta})$ $\mathcal{O}\left(MN^2\right)$}
            \State \text{Evaluate $\mathbf{J}_{2,p} = \mathbf{J}_{1,p}\mathbf{T}(\boldsymbol{\eta})$ $\mathcal{O}\left(MN^2\right)$}
             \EndFor
            
            \For{i = 1 \textbf{to} I}
                \State \text{Evaluate $\mathbf{h}^{(i)}_T(\boldsymbol{\eta}) = \mathbf{J}_0 \mathbf{b}_i$ $\mathcal{O}\left(MN\right)$}
                
                \For{p = 1 \textbf{to} P}
                    \State \text{Evaluate $\mathbf{j}_{2,p,i} = \mathbf{J}_{2,p}\mathbf{b}_i$ $\mathcal{O}\left(MN\right)$}
                    \State \text{Evaluate $\mathbf{j}_{3,p,i} = \mathbf{x}^H_{i}\mathbf{j}_{2,p,i}$ $\mathcal{O}\left(M\right)$}
                    \State \text{Evaluate $\mathbf{j}_{4,p,i} = \left(\mathbf{h}^{(i)}_T(\boldsymbol{\eta})\right)^H\mathbf{j}_{2,p,i}$ $\mathcal{O}\left(M\right)$}
                    \State \text{$d^{(i)}_p(\boldsymbol{\eta}) = -\mathbf{j}_{3,p,i}$}
                    \State \text{$f^{(i)}_p(\boldsymbol{\eta}) = -2\Re\left\{\mathbf{j}_{4,p,i}\right\}$}
   
                \EndFor
        \EndFor
    \end{algorithmic}
\end{algorithm}
In the algorithm we have indicated the complexity for each evaluation. To this regard, we have considered that the inverse of a square matrix entails a complexity equal to the cube of the dimension. Furthermore, to evaluate the complexity of the matrix products we have assumed that the product of an $n \times p$ matrix by a $p \times q$ matrix requires a number of operations proportional to $npq$, neglecting potential optimizations from specialized algorithms for matrix multiplication. Consequently, the overall complexity, defined as $\mathcal{C}_{ECO}$, is:
\begin{equation}
\label{eq:Complexity_ECO}
\begin{aligned}
\mathcal{C}_{ECO} & = \mathcal{O}\left(N^3\right) + \mathcal{O}\left((2P+1)MN^2\right)+\mathcal{O}(IPMN)  \\ & + \mathcal{O}(2IPM)+\mathcal{O}(IMN).
\end{aligned}
\end{equation}
Note that the complexity per iteration depends both on the number of ports $N$ of the ECO, on the number of outputs $M$, on the number of inputs $I$ and on the number of tunable parameters $P$, which can range from $N$, e.g., when the ECO is a diagonal RIS to $N^2$ for fully connected ECOs.

\section{SIM Model}\label{SIM_model}
\begin{figure*}
	\centering
	\includegraphics[width=0.7\textwidth]{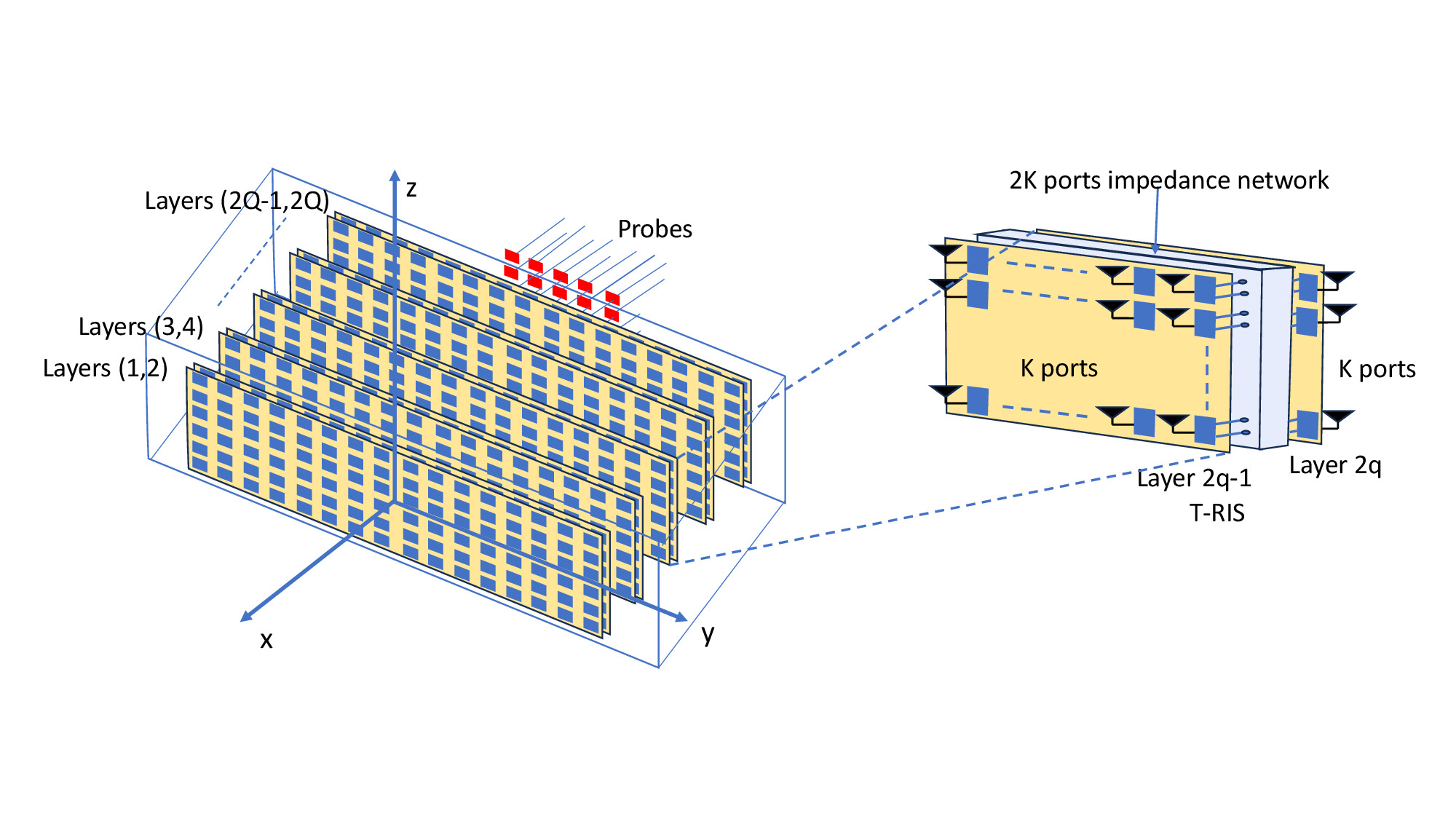}
	\caption{\REV{SIM architecture.}}
	\label{SIM_arch}
\end{figure*}

\begin{figure*}[]
 \centering
     \subfloat[][Pair of facing layers (T-RIS).]{\includegraphics[trim=280 0 280 0,clip,width=0.65\columnwidth]{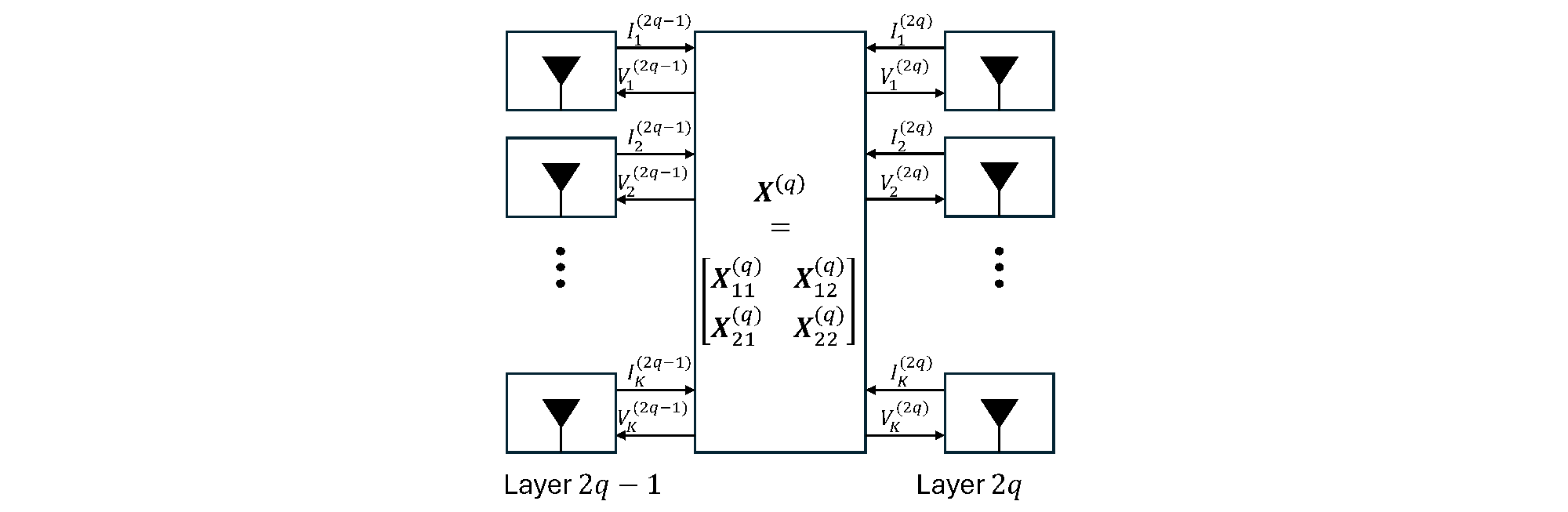} \label{SubFig:SingleLayer}}\quad\quad
     \subfloat[][Overall model.]{\includegraphics[trim=125 0 125 0,clip,width=1.27\columnwidth]{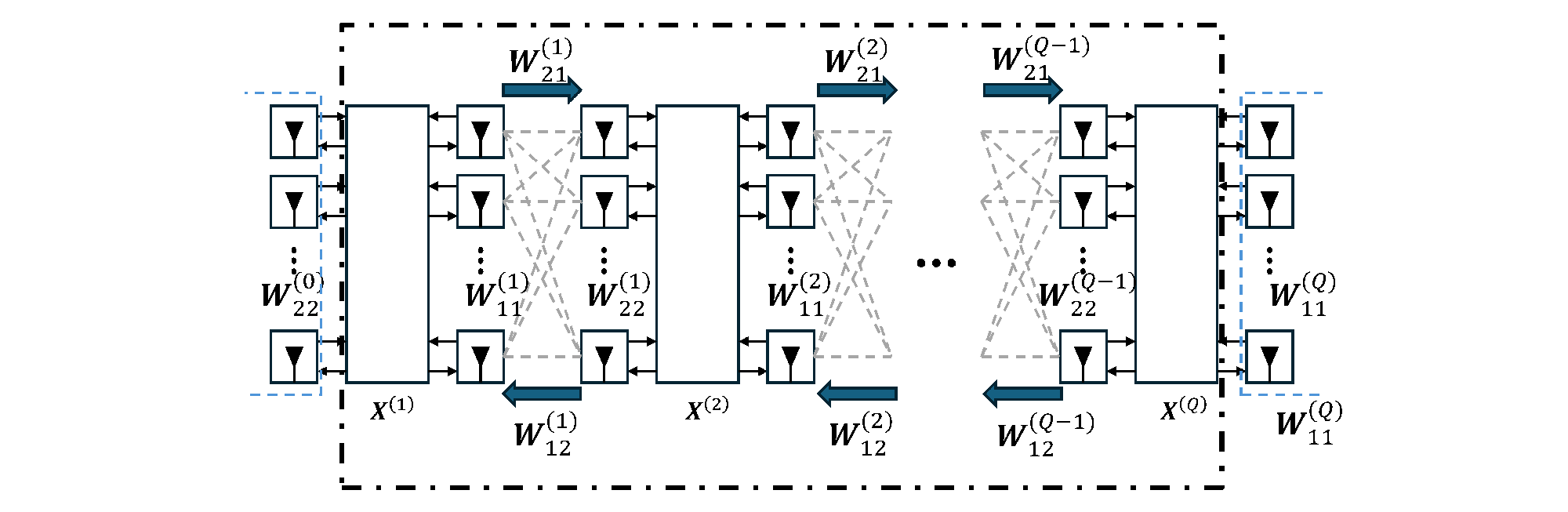}
     \label{SubFig:SimAllLayers}}
     \caption{SIM model.}
     \label{fig:SIM}
\end{figure*}
We now consider the specific case of ECO represented by a SIM. A SIM is a structure housed within a supporting framework that is surrounded by wave-absorbing materials to minimize interference from unwanted diffraction, scattering, and environmental noise \cite{DBLP:journals/ojcs/HassanARDY24,DBLP:journals/jsac/AnXNAHYH23, DBLP:journals/wc/AnYXLNRDH24}. The architecture of a SIM is reported schematically in Fig. \ref{SIM_arch}. Specifically, the SIM is composed of $Q$ couples of facing layers, i.e, with a total of $2Q$ layers. Each couple of facing layers is  a T-RIS modeled as a $2K$ port network. For simplicity, we assume that all layers are characterized by the same number of ports, but the following discussion can be easily generalized to the case where each layer has different dimensions. In this setting, the first layer receives the signal from the external environment, the second layer is connected to the first layer through an internal network, the third layer is connected to the second layer through the wireless channel, and so on, up to the last layer, which is connected to the external environment. In practice, each inner even layer, i.e., for $l = 2, 4, \ldots, 2Q$, receives the signal from the previous layer through an internal network while propagating it to the next layer through the wireless channel. In this scenario, the general model considered previously remains valid with a total number of ports $N = 2QK$. The multi-port model that includes the notations used in the following is shown in Fig. \ref{fig:SIM}.
Note that in \cite{Nerini_Clerckx_SIM}, a multiport S-parameter model for the SIM is proposed, which, due to the equivalence between the $\mathbf{S}$ and $\mathbf{Z}$ matrices, is equivalent to the one presented here. In fact, an equivalent model of the SIM using S-parameters can be obtained thanks to the one-to-one relationship between the $\mathbf{S}$ and $\mathbf{Z}$ matrices (see Eq. (4.44) and Eq. (4.45) in \cite{pozar2011microwave}). However, in our analysis, we chose to adopt the Z-parameter representation, as it allows for a more straightforward handling of situations where there is no direct connection between the ports, such as in wireless channels in the absence of line of sight (LOS) and within the internal network of the SIM. This choice facilitates the derivation of the gradient with respect to the parameters that need to be optimized, thereby enabling the development of an optimization framework, which is the main goal of this work. 

To elaborate, it is worth noting that in a SIM, each layer is only connected to two neighboring layers. As a result, the matrices $\mathbf{Z}_{RE}$, $\mathbf{Z}_{ET}$, $\mathbf{Z}_{EE}$, and $\mathbf{Z}_{E}$ results to be very sparse. Specifically, for $\mathbf{Z}_{ET} \in \mathbb{C}^{2QK \times L}$, only the first $K$ rows are non-zero due to the fact that only the first layer is connected to the external environment. Similarly, for $\mathbf{Z}_{RE} \in \mathbb{C}^{K \times 2QK}$, only the last $K$ columns are non-zero. We are in particular interested in the part of the transfer function in \eqref{eq:TFZ} that contains the effect of the SIM, namely $\mathbf{Z}_{RE} (\mathbf{Z}_{EE}+\mathbf{Z}_{E})^{-1}\mathbf{Z}_{ET}$. For convenience, we consider the matrix $\mathbf{T} = (\mathbf{Z}_{EE}+\mathbf{Z}_{E})^{-1}$ as expressed by $2Q\times 2Q$ sub-matrices $\mathbf{T}_{i,j} \in \mathbb{C}^{K \times K}$, with $i = 1, \ldots, 2Q$, $k = 1,\ldots,2Q$, i.e.:
\begin{equation}\label{TT}
\mathbf{T} = \begin{bmatrix}
\mathbf{T}_{1,1} & \mathbf{T}_{1,2} & \cdots \mathbf{T}_{1,2Q} \\
\mathbf{T}_{2,1} & \mathbf{T}_{2,2} & \cdots \mathbf{T}_{2,2Q}\\
\vdots & \vdots & \vdots  \\
\mathbf{T}_{2Q,1} & \mathbf{T}_{2Q,2} & \cdots \mathbf{T}_{2Q,2Q}\\
\end{bmatrix}.
\end{equation}
Thus, if we denote by $\mathbf{Z}'_{ET} \in \mathbb{C}^{K \times L}$ the matrix composed of the first $K$ rows of $\mathbf{Z}_{ET}$, and  $\mathbf{Z}'_{RE} \in \mathbb{C}^{M \times K}$, the matrix that contains the last $K$ columns of $\mathbf{Z}_{RE}$, we obtain the transfer function $\mathbf{{H}}_{Z}$ as: 
  \begin{equation}\label{eq:TFZSIM}
    \mathbf{{H}}_{Z} = \frac{1}{4Z_0}\left[\mathbf{Z}_{RT}-\mathbf{Z}'_{RE} \mathbf{T}_{2Q,1} \mathbf{Z}'_{ET}\right].
  \end{equation}
Eventually, for the SIM the transfer function in \eqref{eqdef2} can be written as: 
\begin{equation}
\label{eqdef3}
\begin{aligned}
\mathbf{h}^{(i)}_T(\boldsymbol{\eta}) = \mathbf{A} \mathbf{T}_{2Q,1}(\boldsymbol{\eta}) \mathbf{b}_i,
\end{aligned}
\end{equation}
where $\mathbf{A} \in \mathbb{C}^{M \times K}$ and $\mathbf{b}^{(i)} \in \mathbb{C}^{K \times 1}$. Regarding the matrices $\mathbf{Z}_{EE}$ and $\mathbf{Z}_{E}$, they are band matrices that are best decomposed into sub-matrices of $K \times K$ elements. Specifically, as depicted in Fig. \ref{fig:SIM}, we introduce $\mathbf{W}^{(q)}_{i,j} \in \mathbb{C}^{K \times K}$, with $q = 1,2,\ldots,Q-1$, $i = 1,2$, $j = 1,2$, representing the 4 sub-matrices that characterize the connection through the wireless channel between the even layer $2q$ and the odd layer $2q+1$. Additionally, the sub-matrix that characterizes the ports of the first layer of the SIM is denoted as $\mathbf{W}^{(0)}_{2,2} \in \mathbb{C}^{K \times K}$, and the sub-matrix that characterizes the ports of the last layer of the SIM is denoted as $\mathbf{W}^{(Q)}_{1,1} \in \mathbb{C}^{K \times K}$. The matrix $\mathbf{Z}_{EE}$ can be graphically represented as:
\begin{equation}\label{Z_SS}
\mathbf{Z}_{EE} = \begin{bmatrix}
\mathbf{W}^{(0)}_{2,2} & \mathbf{0} & \mathbf{0} & \mathbf{0} & \mathbf{0} \cdots  & \mathbf{0} \\
\mathbf{0} & \mathbf{W}^{(1)}_{1,1} & \mathbf{W}^{(1)}_{1,2} & \mathbf{0} & \mathbf{0} \cdots  & \mathbf{0}\\
\mathbf{0} & \mathbf{W}^{(1)}_{2,1} & \mathbf{W}^{(1)}_{2,2} & \mathbf{0} & \mathbf{0} \cdots  & \mathbf{0} \\
\mathbf{0} & \mathbf{0} & \mathbf{0} & \mathbf{W}^{(2)}_{1,1} & \mathbf{W}^{(2)}_{1,2} \cdots & \mathbf{0} \\
\mathbf{0} & \mathbf{0} & \mathbf{0} & \mathbf{W}^{(2)}_{2,1} & \mathbf{W}^{(2)}_{2,2} \cdots & \mathbf{0} \\
\vdots & \vdots & \vdots & \vdots & \vdots & \vdots \\
\mathbf{0} & \mathbf{0} & \mathbf{0} & \mathbf{0} & \cdots & \mathbf{W}^{(Q)}_{1,1}\\
\end{bmatrix}.
\end{equation}
Regarding the martrix $\mathbf{Z}_E(\boldsymbol{\eta})$, it represents the load network that in a SIM can be seen as $Q$ separate load networks, one for each layer of the SIM. Hence, the controllable parameters independently control each layer. If we define $P_q$ as the number of controllable parameters of layer $q$, with $\sum_q P_q = P$, the vector $\boldsymbol{\eta}$ can be appropriately written as $\boldsymbol{\eta} = \left\{\boldsymbol{\eta}_1,\ldots,\boldsymbol{\eta}_Q\right\}$, where $\boldsymbol{\eta}_q =  \left\{\eta_{q,1},\ldots,\eta_{q,P_q}\right\}$.
We can then introduce the matrices $\mathbf{X}^{(q)}_{i,j}(\boldsymbol{\eta}_q) \in \mathbb{C}^{K \times K}$, with $q = 1,2,\ldots,Q$, $i = 1,2$, $j = 1,2$, which represent the four Z matrices of the connection in the load network between layer $2q-1$ and layer $2q$. Omitting the dependence of $\mathbf{X}^{(q)}_{i,j}$ on $\boldsymbol{\eta}_q$ for ease of representation, the matrix $\mathbf{Z}_E(\boldsymbol{\eta})$ can thus be written as:
\begin{equation}\label{Z_S}
\mathbf{Z}_{E}(\boldsymbol{\eta}) = \begin{bmatrix}
\mathbf{X}^{(1)}_{1,1} & \mathbf{X}^{(1)}_{1,2} & \mathbf{0} & \mathbf{0} \cdots  & \mathbf{0} & \mathbf{0} \\
\mathbf{X}^{(1)}_{2,1} & \mathbf{X}^{(1)}_{2,2} & \mathbf{0} & \mathbf{0} \cdots  & \mathbf{0} & \mathbf{0} \\
\mathbf{0} & \mathbf{0} & \mathbf{X}^{(2)}_{1,1} & \mathbf{X}^{(2)}_{1,2} \cdots & \mathbf{0} & \mathbf{0} \\
\mathbf{0} & \mathbf{0} & \mathbf{X}^{(2)}_{2,1} & \mathbf{X}^{(2)}_{2,2} \cdots & \mathbf{0} & \mathbf{0} \\
\vdots & \vdots & \vdots & \vdots & \vdots & \vdots \\
\mathbf{0} & \mathbf{0} & \mathbf{0} &  \mathbf{0} & \mathbf{X}^{(Q)}_{1,1} & \mathbf{X}^{(Q)}_{1,2}\\
\mathbf{0} & \mathbf{0} & \mathbf{0} & \mathbf{0} & \mathbf{X}^{(Q)}_{2,1} & \mathbf{X}^{(Q)}_{2,2}\\
\end{bmatrix}.
\end{equation}
\subsection{Gradients evaluation for a SIM}
Based on the above, the expression of the gradients given in \eqref{eq:Grad1} can be significantly simplified. To elaborate, let us denote by $\mathbf{Z}_E^{(q)}(\boldsymbol{\eta}_q) = \left\{\mathbf{X}^{(q)}_{i,j}\right\} \in \mathbb{C}^{2K \times 2K}$ the matrix containing the $q$-th block of $\mathbf{Z}_E(\boldsymbol{\eta})$ and by $\mathbf{G}_{q,p}(\boldsymbol{\eta}_q) = \frac{\partial \mathbf{Z}_E^{(q)}(\boldsymbol{\eta}_q)}{\partial \eta_{q,p}} \in \mathbb{C}^{2K \times 2K}$ the tangent matrix of $\mathbf{Z}_E^{(q)}(\boldsymbol{\eta}_q)$ with respect to $\eta_{q,p}$. Then, we introduce:
\begin{equation}
\label{eq:Grad2pre}
\begin{aligned}
d^{(i)}_{q,p}(\boldsymbol{\eta}) & = \frac{\partial ~\mathbf{x}_i^H\mathbf{h}^{(i)}_T(\boldsymbol{\eta})}{\partial \eta_{q,p}} \\
f^{(i)}_{q,p}(\boldsymbol{\eta}) & = \frac{\partial ~\left(\mathbf{h}^{(i)}_T(\boldsymbol{\eta})\right)^H \mathbf{h}^{(i)}_T(\boldsymbol{\eta})}{\partial \eta_{q,p}}.
 \end{aligned}
\end{equation}
Hence, denoting by $\mathbf{R}_q(\boldsymbol{\eta}) = \left\{\mathbf{T}_{2Q,2q-1}(\boldsymbol{\eta}),\mathbf{T}_{2Q,2q}(\boldsymbol{\eta})\right\} \in \mathbb{C}^{K \times 2K}$, and $\mathbf{S}_q(\boldsymbol{\eta}) = \left\{\mathbf{T}^T_{2q-1,1}(\boldsymbol{\eta}),\mathbf{T}_{2q,1}^T(\boldsymbol{\eta})\right\}^T \in \mathbb{C}^{2K \times K}$, it is straightforward to get:
\begin{equation}
\label{eq:Grad2}
\begin{aligned}
d^{(i)}_{q,p}(\boldsymbol{\eta}) & =  -\mathbf{x}_i^H\mathbf{A}\mathbf{R}_q(\boldsymbol{\eta})\mathbf{G}_{q,p}(\boldsymbol{\eta}_q)\mathbf{S}_q(\boldsymbol{\eta})\mathbf{b}_i\\
f^{(i)}_{q,p}(\boldsymbol{\eta}) & =  -2\Re\left\{\left(\mathbf{h}^{(i)}_T(\boldsymbol{\eta})\right)^H\mathbf{A}\mathbf{R}_q(\boldsymbol{\eta})\mathbf{G}_{q,p}(\boldsymbol{\eta}_q)\mathbf{S}_q(\boldsymbol{\eta})\mathbf{b}_i\right\}.
\end{aligned}
\end{equation}
Given the band structure of the matrices $\mathbf{Z}_{EE}$ and $\mathbf{Z}_{E}$, it is possible to derive an iterative approach for the calculation of $\mathbf{R}_q(\boldsymbol{\eta})$ and $\mathbf{S}_q(\boldsymbol{\eta})$ that does not require evaluating the inverse matrix $\mathbf{T}(\boldsymbol{\eta})$, which has a complexity of $\mathcal{O}(N^3)$. 
\REV{The details of the iterative algorithm that significantly reduces complexity are provided in Appendix A. Thanks to this algorithm, it is easy to see that the complexity of calculating} $\mathbf{R}_q(\boldsymbol{\eta})$ and $\mathbf{S}_q(\boldsymbol{\eta})$ depends on the computation of products and inverses of matrices of size $K \times K$ for a number of times proportional to the number of levels $Q$. More specifically, the calculation of each term $\mathbf{M}_r$, $r = 2,\ldots,2Q$ requires 3 products and an inversion of $K \times K$ matrices, resulting in an overall complexity of $\mathcal{O}\left(4(2Q-1)K^3\right)$. By similar reasoning, it can be seen that the calculation of $\mathbf{U}_r$ requires the same complexity as $\mathbf{M}_r$, while the $\mathbf{T}_r$ require a complexity of $\mathcal{O}\left(2Q + 2)K^3\right)$. Therefore, the overall complexity for the computation of $\mathbf{R}_q(\boldsymbol{\eta})$ and $\mathbf{S}_q(\boldsymbol{\eta})$ is approximately $\mathcal{O}\left(18QK^3\right)$.

Thus, the terms $d^{(i)}_p(\boldsymbol{\eta})$ and $f^{(i)}_p(\boldsymbol{\eta})$ in \eqref{eq:Grad2} can be evaluated following algorithm \ref{Algo2} for an overall complexity:
\begin{equation}
\label{eq:Complexity_SIM}
\begin{aligned}
\mathcal{C}_{SIM} & = \mathcal{O}\left(18QK^3\right) + \mathcal{O}\left(2PMK^2\right)+\mathcal{O}\left(QMK^2\right) \\ & + \mathcal{O}(IPMK) + \mathcal{O}(2IPM)+\mathcal{O}(IQMK),
\end{aligned}
\end{equation}
where in \eqref{eq:Complexity_SIM} we used the fact that $P = \sum\limits_{q=1}^{Q} P_q$. 
\begin{algorithm}
    \caption{$SIM$: Evaluation of 
$d^{(i)}_{q,p}(\boldsymbol{\eta})$ and $f^{(i)}_{q,p}(\boldsymbol{\eta})$}\label{Algo2}
    \small
    \begin{algorithmic}[1]
        \State \textbf{Input:} $\boldsymbol{\eta}, \mathbf{A}, \mathbf{b}_i, \mathbf{x}_i, \mathbf{Z}_{EE},  \mathbf{Z}_{E}(\boldsymbol{\eta}), \mathbf{G}_{p,q}(\boldsymbol{\eta}_q)$
        
        \State \text{Evaluate $\mathbf{R}_q(\boldsymbol{\eta})$ and $\mathbf{S}_q(\boldsymbol{\eta})$ according to \eqref{eq:R1it_0}-\eqref{eq:C1it_3} $\mathcal{O}\left(18QK^3\right)$}
        \For{q = 1 \textbf{to} Q}
        \State \text{Evaluate $\mathbf{J}_{0,q} = \mathbf{A}\mathbf{R}_q(\boldsymbol{\eta})$ $\mathcal{O}\left(MK^2\right)$}
        
        \For{p = 1 \textbf{to} $P_q$}
            \State \text{Evaluate $\mathbf{J}_{1,q,p} = \mathbf{J}_{0,q}\mathbf{G}_{q,p}(\boldsymbol{\eta}_q)$ $\mathcal{O}\left(MK^2\right)$}
            \State \text{Evaluate $\mathbf{J}_{2,q,p} = \mathbf{J}_{1,q,p}\mathbf{S}_q(\boldsymbol{\eta})$ $\mathcal{O}\left(MK^2\right)$}
             \EndFor
            
            \For{i = 1 \textbf{to} I}
                \State \text{Evaluate $\mathbf{h}^{(i)}_T(\boldsymbol{\eta}) = \mathbf{J}_{0,q} \mathbf{b}_i$ $\mathcal{O}\left(MK\right)$}
                
                \For{p = 1 \textbf{to} $P_q$}
                    \State \text{Evaluate $\mathbf{j}_{2,q,p,i} = \mathbf{J}_{2,q,p}\mathbf{b}_i$ $\mathcal{O}\left(MK\right)$}
                    \State \text{Evaluate $\mathbf{j}_{3,q,p,i} = \mathbf{x}^H_{i}\mathbf{j}_{2,q,p,i}$ $\mathcal{O}\left(M\right)$}
                    \State \text{Evaluate $\mathbf{j}_{4,q,p,i} = \left(\mathbf{h}^{(i)}_T(\boldsymbol{\eta})\right)^H\mathbf{j}_{2,q,p,i}$ $\mathcal{O}\left(M\right)$}
                    \State \text{$d^{(i)}_{q,p}(\boldsymbol{\eta}) = -\mathbf{j}_{3,q,p,i}$}
                    \State \text{$f^{(i)}_{q,p}(\boldsymbol{\eta}) = -2\Re\left\{\mathbf{j}_{4,q,p,i}\right\}$}
   
                \EndFor
        \EndFor
        \EndFor
    \end{algorithmic}
\end{algorithm}

\subsection{SIM with diagonal T-RISs}
When the faced layers of the SIM are composed of diagonal T-RISs \cite{Nerini_Clerckx_SIM},
each load network element of layer $2q-1$ of the SIM is connected to a single element of layer $2q$, with $q = 1,\ldots,Q$. Hence, the load network can be decomposed into $K$ two-ports networks $\mathbf{D}^{(q)}_k \in \mathbb{C}^{2 \times 2}$ with elements $D^{(q)}_{k}(n,m)$ for $n = 1,2$ and $m = 1,2$. In this setting, the matrices $\mathbf{X}^{(q)}_{n,m}$ in \eqref{Z_S} are shown to be diagonal matrices containing the element $D^{(q)}_{k}(n,m)$ in the $k$-th diagonal entry:
\begin{equation}\label{X_diag}\small
\mathbf{X}^{(q)}_{n,m} = 
\begin{bmatrix}
D_1^{(q)}(n,m) & 0 & \cdots & 0 \\
0 & D_2^{(q)}(n,m) & \cdots & 0 \\
\vdots & \vdots & \ddots & \vdots \\
0 & 0 & \cdots & D_K^{(q)}(n,m)
\end{bmatrix}.    
\end{equation}
Since in a SIM each layer must operate in transmissive mode, the two-port network $D^{(q)}_{p}(n,m)$, $p = 1,\ldots,K$, can be characterized by a single tunable parameter $\eta_{q,p}$, representing the transmission coefficient angle \cite{Nerini_Clerckx_SIM}, i.e., in this case $P_q = K$ and $P = QK$. In the S-parameter representation, a two-port network of this type has diagonal elements $S_{1,1}$ and $S_{2,2}$ equal to zero, while $S_{2,1} = S_{1,2} = e^{j\eta_{q,p}}$. In the Z-parameter representation, we then have \cite{pozar2011microwave}:
\begin{equation}\label{D22_mat}
\mathbf{D}^{(q)}_{p} = j Z_0 \begin{bmatrix}
\frac{\cos(\eta_{q,p})}{\sin(\eta_{q,p})} &  \frac{1}{\sin(\eta_{q,p})} \\
\frac{1}{\sin(\eta_{q,p})}  & \frac{\cos(\eta_{q,p})}{\sin(\eta_{q,p})}
\end{bmatrix}.
\end{equation}
Denoting $\mathbf{D}'^{(q)}_{p} = \frac{\partial \mathbf{D}^{(q)}_{p}}{\partial \eta_{q,p}} \in \mathbb{C}^{2 \times 2}$, we have:
\begin{equation}\label{D22_mat_grad}
\mathbf{D}'^{(q)}_{p} = -j Z_0 \begin{bmatrix}
\left(\frac{\cos(\eta_{q,p})}{\sin(\eta_{q,p})}\right)^2+1 &  \frac{\cos(\eta_{q,p})}
{\sin^2(\eta_{q,p})} \\ \frac{\cos(\eta_{q,p})}
{\sin^2(\eta_{q,p})}  & \left(\frac{\cos(\eta_{q,p})}{\sin(\eta_{q,p})}\right)^2+1
\end{bmatrix}.
\end{equation}
It is then straightforward to observe that $\mathbf{G}_{p,q}(\boldsymbol{\eta}_q) = \mathbf{G}_{p,q}({\eta}_{q,p})$, i.e., $\mathbf{G}_{q,p}$ is a function of ${\eta}_{q,p}$ only. Hence, if we introduce the element-selection diagonal matrices $\mathbf{J}_p \in \mathbb{C}^{K \times K}$ consisting of all zeros except in the p-th diagonal element, which is one, we easily get:
\begin{equation}\label{D22_mat_grad_2}
\begin{aligned}
    \mathbf{G}_{p,q}({\eta}_{q,p}) \begin{bmatrix}
D'^{(q)}_p(1,1)\mathbf{J}_p & D'^{(q)}_p(1,2)\mathbf{J}_p \\ D'^{(q)}_p(2,1)\mathbf{J}_p  & D'^{(q)}_p(2,2)\mathbf{J}_p
\end{bmatrix}.
\end{aligned}
\end{equation}
Therefore, the expression of the gradients in $\eqref{eq:Grad2}$ can also be simplified. To this end, we introduce the vectors $\mathbf{t}^{(1)}_{q,p}(\boldsymbol{\eta}) \in \mathbb{C}^{K \times 1}$ and $\mathbf{t}^{(2)}_{q,p}(\boldsymbol{\eta}) \in \mathbb{C}^{K \times 1}$ as the $p$-th columns of the matrices $\mathbf{T}_{2Q,2q-1}(\boldsymbol{\eta})$ and $\mathbf{T}_{2Q,2q}(\boldsymbol{\eta})$, respectively. Similarly, we introduce the vectors $\mathbf{t}^{(3)}_{q,p}(\boldsymbol{\eta}) \in \mathbb{C}^{1 \times K}$ and $\mathbf{t}^{(4)}_{q,p}(\boldsymbol{\eta}) \in \mathbb{C}^{1 \times K}$ as the $p$-th rows of the matrices $\mathbf{T}_{2q-1,1}(\boldsymbol{\eta})$ and $\mathbf{T}_{2q,1}(\boldsymbol{\eta})$, respectively. Therefore, introducing $\mathbf{F}_{q,p}$ defined in \eqref{eq:Grad3_pre}, we get:
\begin{figure*}[] 
\centering
\begin{equation}
\label{eq:Grad3_pre}
\begin{aligned}
\mathbf{F}_{q,p}(\boldsymbol{\eta}) = \left(D'^{(q)}_p(1,1) \mathbf{t}^{(1)}_{q,p}(\boldsymbol{\eta}) + D'^{(q)}_p(2,1)\mathbf{t}^{(2)}_{q,p}(\boldsymbol{\eta})\right)\mathbf{t}^{(3)}_{q,p}(\boldsymbol{\eta}) + \left(D'^{(q)}_p(1,2) \mathbf{t}^{(1)}_{q,p}(\boldsymbol{\eta}) + D'^{(q)}_p(2,2) \mathbf{t}^{(2)}_{q,p}(\boldsymbol{\eta})\right)\mathbf{t}^{(4)}_{q,p}(\boldsymbol{\eta}).
\end{aligned}
\end{equation}
\end{figure*}
\begin{equation}
\label{eq:Grad3}
\begin{aligned}
d^{(i)}_{q,p}(\boldsymbol{\eta}) & =  -\mathbf{x}_i^H\mathbf{A}\mathbf{F}_{q,p}(\boldsymbol{\eta})\mathbf{b}_i\\
f^{(i)}_{q,p}(\boldsymbol{\eta}) & =  -2\Re\left\{\left(\mathbf{h}^{(i)}_T(\boldsymbol{\eta})\right)^H\mathbf{A}\mathbf{F}_{q,p}(\boldsymbol{\eta})\mathbf{b}_i\right\}.
\end{aligned}
\end{equation}
As for the evaluation of $\mathbf{F}_{q,p}$ in \eqref{eq:Grad3_pre}, it entails 4 products of $K$ dimensional vectors for each $q$ and each $p$, yielding a complexity $\mathcal{O}\left(4KP\right)$. The procedure for computing $d^{(i)}_{q,p}$ and $f^{(i)}_{q,p}$ from equation \eqref{eq:Grad3} is reported in algorithm \ref{Algo3}. The complexity for the diagonal SIM, denotes $D-SIM$ is then:
\begin{equation}
\label{eq:Complexity_SIMd}
\begin{aligned}
\mathcal{C}_{D-SIM} & = \mathcal{O}\left(18QK^3\right) + \mathcal{O}\left(2PMK^2\right) + \mathcal{O}(IPMK)  \\ & + \mathcal{O}(2IPM)+\mathcal{O}(IQMK) + \mathcal{O}(4PK).
\end{aligned}
\end{equation}

\begin{algorithm}
    \caption{$D-SIM$: Evaluation of 
$d^{(i)}_{q,p}(\boldsymbol{\eta})$ and $f^{(i)}_{q,p}(\boldsymbol{\eta})$}\label{Algo3}
    \small
    \begin{algorithmic}[1]
        \State \textbf{Input:} $\boldsymbol{\eta}, \mathbf{A}, \mathbf{b}_i, \mathbf{x}_i, \mathbf{Z}_{EE},  \mathbf{Z}_{E}(\boldsymbol{\eta}), \mathbf{G}_{p,q}(\boldsymbol{\eta}_q)$
        
        \State \text{Evaluate $\mathbf{R}_q(\boldsymbol{\eta})$ and $\mathbf{S}_q(\boldsymbol{\eta})$ according to \eqref{eq:R1it_0}-\eqref{eq:C1it_3} $\mathcal{O}\left(18QK^3\right)$}
        \For{q = 1 \textbf{to} Q}
        
        \For{p = 1 \textbf{to} $P_q$}
            \State \text{Evaluate $\mathbf{F}_{q,p}(\boldsymbol{\eta})$ according to \eqref{eq:Grad3_pre} $\mathcal{O}\left(4K\right)$}
            \State \text{Evaluate $\mathbf{J}_{1,q,p} = \mathbf{A}\mathbf{F}_{q,p}(\boldsymbol{\eta})$ $\mathcal{O}\left(MK^2\right)$}
             \EndFor
            
            \For{i = 1 \textbf{to} I}
                \State \text{Evaluate $\mathbf{h}^{(i)}_T(\boldsymbol{\eta}) = \mathbf{J}_{0,q} \mathbf{b}_i$ $\mathcal{O}\left(MK\right)$}
                
                \For{p = 1 \textbf{to} $P_q$}
                    \State \text{Evaluate $\mathbf{j}_{2,q,p,i} = \mathbf{J}_{1,q,p}\mathbf{b}_i$ $\mathcal{O}\left(MK\right)$}
                    \State \text{Evaluate $\mathbf{j}_{3,q,p,i} = \mathbf{x}^H_{i}\mathbf{j}_{2,q,p,i}$ $\mathcal{O}\left(M\right)$}
                    \State \text{Evaluate $\mathbf{j}_{4,q,p,i} = \left(\mathbf{h}^{(i)}_T(\boldsymbol{\eta})\right)^H\mathbf{j}_{2,q,p,i}$ $\mathcal{O}\left(M\right)$}
                    \State \text{$d^{(i)}_{q,p}(\boldsymbol{\eta}) = -\mathbf{j}_{3,q,p,i}$}
                    \State \text{$f^{(i)}_{q,p}(\boldsymbol{\eta}) = -2\Re\left\{\mathbf{j}_{4,q,p,i}\right\}$}
   
                \EndFor
        \EndFor
        \EndFor
    \end{algorithmic}
\end{algorithm}
Note that the complexity obtained above is slightly overestimated because it does not take into account that the diagonal nature of the matrices $\mathbf{X}^{(q)}_{n,m}$ reduces the complexity in the calculation of $\mathbf{R}_q(\boldsymbol{\eta})$ and $\mathbf{S}_q(\boldsymbol{\eta})$. Moreover, it is worth noting that the reduction in complexity due the diagonal case stems mainly from the reduction of the number of variables $P$, which in the diagonal case is $P = QK$, whereas in the non diagonal or beyond diagonal case, it is $P \ge QK$.  
\subsection{Unilateral approximation}
In \cite{Nerini_Clerckx_SIM}, it is shown that the model used in all the works addressing SIM so far relies on various approximations, including the unilateral approximation. Essentially, this approximation consists of assuming that the interaction between layers $2q$ and $2q+1$ of the SIM occurs in one direction only, meaning that the wireless channel separating two SIMs is not reciprocal. In this case, we have $\mathbf{W}^{(q)}_{1,2} = \mathbf{0}$, $\forall q$, in the expression of $\mathbf{Z}_{EE}$ reported in \eqref{Z_SS}. In this case, it is possible to simplify the iterative procedure for calculating the transfer function $\mathbf{T}_1 = \mathbf{T}_{2Q,1}(\boldsymbol{\eta})$. \REV{The details of this derivation are reported in Appendix B.} 
The iterative procedure in this case is slightly simplified compared to the general $SIM$ case, although the order of magnitude of the complexity remains the same. In particular, it is easy to see that the complexity of the iterative process is reduced by a factor of 3, resulting in an overall complexity for the $SIM$ case with unilateral approximation, denoted as $\mathcal{C}_{U-SIM}$, equal to:
\begin{equation}
\label{eq:Complexity_SIMu}
\begin{aligned}
\mathcal{C}_{U-SIM} & = \mathcal{O}\left(6QK^3\right)  + \mathcal{O}\left(2PMK^2\right)+\mathcal{O}\left(QMK^2\right) \\ & + \mathcal{O}(IPMK)  + \mathcal{O}(2IPM)+\mathcal{O}(IQMK).
\end{aligned}
\end{equation}
Moreover, denoting by $\mathcal{C}_{DU-SIM}$ the complexity in the case of unilateral approximation with diagonal T-RIS, denoted as $DU-SIM$, we have:
\begin{equation}
\label{eq:Complexity_SIMdu}
\begin{aligned}
\mathcal{C}_{DU-SIM} & = \mathcal{O}\left(6QK^3\right) + \mathcal{O}\left(2PMK^2\right) + \mathcal{O}(IPMK)  \\ & + \mathcal{O}(2IPM)+\mathcal{O}(IQMK) + \mathcal{O}(4PK).
\end{aligned}
\end{equation}

\subsection{Unilateral approximation with diagonal and ideal T-RISs}
The case of ideal T-RISs refer to the case in which there is no coupling between the elements of the T-RISs constituting the SIM. Moreover, the matrices $\mathbf{W}^{(q)}_{1,1}$ and $\mathbf{W}^{(q-1)}_{2,2}$ are characterized by the impedances at the ports $Z_0$. Specifically, we have $\mathbf{W}^{(q)}_{1,1} = \mathbf{W}^{(q-1)}_{2,2} = Z_0 \mathbf{I}_K$. To elaborate, under the above assumptions the matrices $\boldsymbol{\Omega}_q$ in \eqref{eq:Uni2} are diagonal. Moreover, from \eqref{D22_mat}, and denoting $\eta = \eta_{q,p}$, each diagonal element ${\omega}_{q,p}$ of $\boldsymbol{\Omega}_q$ takes the form:
\begin{equation}
\label{eq:Uni5_ideal}
\begin{aligned}
{\omega}_{q,p} & = \frac{1}{Z_0}\left(1 +j\frac{\cos\eta}{\sin\eta}-j\frac{1}{\sin\eta}\left(1+j\frac{\cos\eta}{\sin\eta}\right)^{-1} j\frac{1}{\sin\eta}\right)^{-1}\\ & =
\frac{1}{Z_0}\left(\frac{\sin\eta+j\cos\eta}{\sin\eta}+\frac{1}{\sin\eta(\sin\eta+j\cos\eta)}\right)^{-1}\\
& = \frac{1}{Z_0}\left(\frac{\sin^2\eta-\cos^2\eta+2j\cos\eta\sin\eta+1}{\sin\eta(\sin\eta+j\cos\eta)}\right)^{-1}\\
& = \frac{1}{2Z_0}.
\end{aligned}
\end{equation}
It is easy to verify with similar steps that $\boldsymbol{\zeta}_{q}$ is diagonal with the $p$-th entry equal to ${\zeta}_{q,p} = \frac{1}{2Z_0}$. Furthermore, the matrix $\mathbf{X}^{(q)}_{2,1}\left(\mathbf{X}^{(q)}_{1,1}+\mathbf{W}^{(q-1)}_{2,2}\right)^{-1}$ that appears in \eqref{eq:Uni3} is also diagonal.  Let us denote this matrix by $\mathbf{Y}_q$ 
with $Y_{q,p}$ representing its $p$-th entry. We have:
\begin{equation}
\label{eq:Uni6_ideal_0}
\begin{aligned}
Y_{q,p} & = j\frac{1}{\sin\eta_{q,p}}\left(1+j\frac{\cos\eta_{q,p}}{\sin\eta_{q,p}}\right)^{-1}\\
& = e^{j\eta_{q,p}}.
\end{aligned}
\end{equation}
For the sake of notation, we denote by $\mathbf{Y}_q = e^{j\boldsymbol{\eta}_{q}}$. From \eqref{eq:Uni5}, \eqref{eq:Uni2}, \eqref{eq:Uni3} and \eqref{eq:Uni4}, setting $\mathbf{W}_{2,1}^{(Q)} = \mathbf{I}_K$, we can derive

\begin{equation}
\label{eq:Uni6_ideal}
\begin{aligned}
\mathbf{T}_{2Q,1} & = \left(-\frac{1}{2Z_0}\right)^Qe^{j\boldsymbol{\eta}_{Q}}\prod\limits_{q=Q-1,Q-2,\ldots,1} \mathbf{W}_{2,1}^{(q)}e^{j\boldsymbol{\eta}_{q}}.
\end{aligned}
\end{equation}

It is noted from \eqref{eqdef3} that \eqref{eq:Uni6_ideal} represents the I/O relationship of the SIM, which becomes a cascade comprising the propagation through the channels that separate the levels $2q$ and $2q+1$, represented by the terms $\frac{1}{2Z_0}\mathbf{W}_{2,1}^{(q)}$, along with the phase shifts introduced during the transition from level $2q-1$ to level $2q$. In this particular case, therefore, the SIM model coincides with that traditionally used in all previous works, e.g., see \cite{DBLP:journals/ojcs/HassanARDY24,DBLP:journals/jsac/AnXNAHYH23, DBLP:journals/wc/AnYXLNRDH24,DBLP:journals/vtm/BasarALWJYDS24,DBLP:conf/icc/AnRDY23,DBLP:conf/icc/AnY0RDPH24,DBLP:journals/jsac/AnYGRDPH24,DBLP:journals/wcl/NiuAPGCD24,SIM_ISAC,AntiJammingSIM}.

It should be noted that in the models used to characterize the SIM thus far, the terms $\frac{1}{2Z_0}\mathbf{W}_{2,1}^{(q)}$ have been modeled using the Rayleigh-Sommerfeld diffraction equation \cite{lin2018all}, which has been applied in the context of all-optical diffractive deep neural networks (D2NN). However, its direct application to SIMs operating at radio frequencies may be questionable and may deserve future studies. With the proposed model with parameters Z, it more generally represents the coupling in terms of voltage to current between the ports of two T-RIS at different levels of the SIM.

\subsubsection{Gradient computation through back propagation}
The gradient of the error function has been calculated in previous works that have used the same SIM model, for example \cite{DBLP:journals/wc/AnYXLNRDH24,DBLP:journals/vtm/BasarALWJYDS24,DBLP:conf/icc/AnRDY23,DBLP:conf/icc/AnY0RDPH24, DBLP:journals/jsac/AnYGRDPH24}. Below, we provide an algorithm for calculating this gradient based on the back-propagation mechanism. To elaborate, considering a generic input/output $\mathbf{b}_i$/$\mathbf{x}_i$, we introduce $\mathbf{B}_{q} \in \mathbb{C}^{K \times 1}$ and $\mathbf{O}_{q} \in \mathbb{C}^{K \times 1}$ as:
\begin{equation}
\label{eq:BP_1} 
\begin{aligned}
\mathbf{v}_{1} & = \mathbf{b}_i &\\
\mathbf{O}_{q} & = -\frac{1}{2Z_0}e^{j\boldsymbol{\eta}_{q}}\mathbf{v}_{q} & \text{for $q = 1,\ldots,Q$}\\
\mathbf{v}_{q} & = \mathbf{W}_{2,1}^{(q)} \mathbf{O}_{q-1} & \text{for $q = 2,\ldots,Q$},\\
\end{aligned}
\end{equation}
so that it is easy to get from \eqref{eq:Uni6_ideal} $\mathbf{T}_{2Q,1} = \mathbf{O}_{Q}$. Accordingly, the error in \eqref{eq:mMSEMat0_1} can be written as:
\begin{equation}
\label{eq:mMSEMat0_1bis}
\begin{aligned}
\epsilon = \left(\mathbf{A}\mathbf{O}_{Q}-{\mathbf{x}_i}\right)^H\left(\mathbf{A}\mathbf{O}_{Q}-\mathbf{x}_i\right).
\end{aligned}
\end{equation}
We have:
\begin{equation}
\label{eq:BP_3}
\begin{aligned}
\boldsymbol{\lambda}_{Q} = \frac{\partial \epsilon}{\partial \mathbf{O}_{Q}} = \mathbf{A}^H\left(\mathbf{A}\mathbf{O}_{Q}-{\mathbf{x}_i}\right).
\end{aligned}
\end{equation}
From the definitions in \eqref{eq:BP_1}, it is easy to find the iterative relationship:
\begin{equation}
\label{eq:BP_4}
\begin{aligned}
\boldsymbol{\lambda}_{q-1} = \frac{\partial \epsilon}{\partial \mathbf{O}_{q-1}} =-\frac{1}{2Z_0}\left[e^{j{\boldsymbol{\eta}}_{q}}\mathbf{W}_{2,1}^{(q)}\right]^H\boldsymbol{\lambda}_{q},
\end{aligned}
\end{equation}
for $q= Q,Q-1,\ldots,2$. Hence, the terms $\boldsymbol{\lambda}_{q}$ can be evaluated iteratively from $\boldsymbol{\lambda}_{Q}$ following the backward propagation algorithm \eqref{eq:BP_4}. 
From \eqref{eq:BP_1} and \eqref{eq:BP_4}, denoting by $\boldsymbol{\mu}_q = \frac{\partial \epsilon}{\partial \boldsymbol{\eta}_{q}}$ we have:
\begin{equation}
\label{eq:BP_6}
\begin{aligned}
\boldsymbol{\mu}_q & = \Re\left[\left(\frac{\partial \mathbf{O}_{q}}{\partial \boldsymbol{\eta}}\right)^H\frac{\partial \epsilon}{\partial \mathbf{O}_{q}}\right] \\ & = -\frac{1}{2Z_0}\Re\left[\left(j e^{j{\boldsymbol{\eta}}_{q}}\mathbf{v}_q\right)^H\text{diag}\left(\boldsymbol{\lambda}_{q}\right)\right].
\end{aligned}
\end{equation}
The calculation of the terms $\boldsymbol{\lambda}_q$ can be performed from $Q$ backward to index $1$ with a complexity of $\mathcal{O}(QK^2)$ after computing the terms in \eqref{eq:BP_1} with forward propagation, which also requires a complexity of $\mathcal{O}(QK^2)$. This approach closely resembles the backpropagation algorithm used in a classic neural networks, although the architecture is quite different here, as the tunable parameters are not in the weights of the channel $\mathbf{W}_{2,1}^{(q)}$, but rather in the phase shifts introduced at each node. In summary, considering that the algorithm must be run for all the $I$ input/output pairs, we arrive at a complexity for the case at hand, called $DU-SIM_{id}$, equal to:
\begin{equation}
\label{eq:Complexity_SIMdu_2}
\begin{aligned}
\mathcal{C}_{DU-SIM_{id}} = \mathcal{O}(2IQK^2).
\end{aligned}
\end{equation}

\section{Simulation Results}
\label{sec:results}
In the following, we will present some results obtained by considering a specific optimization problem, that is the realization of the 2D DFT transfer function, similarly to what was considered in \cite{DBLP:journals/jsac/AnYGRDPH24}.
We aim to demonstrate that the SIM operates as desired even in cases where the ideal assumptions made in previous works—namely, the unilateral assumption and the assumption of the absence of mutual couplings—are not valid, thus necessitating the use of the general SIM model. For this purpose, we will consider the $D-SIM$ architecture for simplicity, leaving the study of more elaborate T-RIS architectures, such as the beyond-diagonal structures presented in {\cite{Nerini_Clerckx_SIM}, for future research.
It will be shown that the $D-SIM$ case outperforms the $DU-SIM_{id}$ case, which has been the model used in previous works.
As an additional goal, we intend to show that if we perform optimization assuming an ideal model $DU-SIM_{id}$ but the intrinsic assumptions of that model are not valid, meaning there is a mismatch between the model used for optimization and the real model, the performance will be, as expected, very poor.

\subsection{Simulation setup}
We assume the carrier frequency $f_0 = 28$ GHz, yielding a wavelength $\lambda = 0.0107$ m.
As for each element of the $q$-th T-RIS of the SIM, with $q = 1,2,\ldots,Q$, we consider dipoles arranged as a uniform planar array (UPA) with $N_y^{(q)}$ elements along the $y$ axis and $N_z^{(q)}$ elements along the $z$ axis.
For the calculation of the matrices $\mathbf{W}^{(q)}_{i,j} \in \mathbb{C}^{K \times K}$, with $q = 1,2,\ldots,Q-1$, we therefore used the analytical procedure proposed in \cite{DR1}, which depends on the arrangement and dimensions of the dipoles.
Specifically, we follow the model proposed in \cite{Abrardo24RisOpt}, in which each element is a metallic dipole with a radius of $\lambda/500$ and a length of $L = 0.46 \lambda$.
\REV{Three different values for the dipole spacing in the $y$ direction are considered: $d_y = \frac{\lambda}{3}$, $d_y = \frac{\lambda}{2}$, and $d_y = \frac{2\lambda}{3}$, corresponding to three different levels of densification of the constituent $T$-RISs. Conversely, in the $z$ direction, we consider a fixed dipole spacing of $d_z = \frac{3}{4} \lambda$, which, for dipoles with length $L = 0.46 \lambda$ oriented in the $z$ direction, can be regarded as the maximum level of packing density. Furthermore, we consider three different values for the spacing $d_x$ between two adjacent layers: $d_x = \frac{\lambda}{2}$, $d_x = \lambda$, and $d_x = 2\lambda$, corresponding to three different thicknesses of the SIM equal to $(Q-1)d_x$, whereas the thickness of each T-RIS is assumed to be negligible.}
The results are obtained by considering the minimization of the error between the target 2D DFT matrix and the EM response of the SIM.
Specifically, we consider that the probes are identical dipoles, like those constituting the SIM, arranged in a UPA with $d_y = \lambda/2$, $d_z = \frac{3}{4} \lambda$, with $L_y$ and $L_z$ elements along the $y$ and $z$ axes, respectively, and at a distance of $\lambda$ from the $Q$-th layer of the SIM.
In other words, the probes are centered with respect to the last layer of the SIM and are arranged as an additional layer of the SIM with $M = L_y L_z$ elements.
Furthermore, each probe is assumed to be match-terminated at $Z_0$ and the $\mathbf{Z}'_{RE}$ matrix is computed following the same approach used to evaluate the matrices $\mathbf{W}^{(q)}_{i,j}$.
Regarding the first layer of the SIM, it contains the same geometry as the probes, with $N_y^{(1)} = L_y$ and $N_z^{(1)} = L_z$ elements, while the other layers are all characterized by the same geometry with $N_y^{(q)} = N_y$ and $N_z^{(q)} = N_z$ elements. 

\subsection{2D DFT approximation through SIM}
To approximate the 2D DFT function, we refer to the general optimization problem formulated in \eqref{eq:Transf2}, in which $\mathbf{A} = \mathbf{Z}'_{RE}$ and $\boldsymbol{\Theta}$ represents the 2D DFT matrix.
To elaborate, let us introduce the row and column indices $m_y$ and $m_z$ for the input layer and $n_y$ and $n_z$  for the output layer, with $n_y, m_y = 1,\ldots,L_y$ and $n_z, m_z = 1,\ldots,L_z$.
Consequently, by denoting $m =  (m_z-1)L_z+m_y$ and $n =  (n_z-1)L_z+n_y$, we can express $\boldsymbol{\Theta} \in \mathcal{C}^{M \times M}$ as:
\begin{equation}
\label{eq:2DFT_MAT}
\begin{aligned}
\left[\boldsymbol{\Theta}\right]_{n,m} = e^{-j2\pi\frac{(m_y-1)(n_y-1)}{L_y}}e^{-j2\pi\frac{(m_z-1)(n_z-1)}{L_z}}.
\end{aligned}
\end{equation}

More specifically, as proposed in \cite{DBLP:journals/jsac/AnYGRDPH24}, we consider a scaling factor $\beta$ that serves as normalization for the error function, thereby redefining the optimization problem as:
\begin{equation}
\label{eq:Transf3}
\begin{aligned}
\min \limits_{\boldsymbol{\eta}} \sum\limits_{i } \left( \beta\mathbf{y}_i(\boldsymbol{\eta})-\mathbf{x}_i\right)^H\left(\beta \mathbf{y}_i(\boldsymbol{\eta})-\mathbf{x}_i\right).
\end{aligned}
\end{equation}
Problem \eqref{eq:Transf3} is solved using the gradient descent mechanism shown in \eqref{eq:mMSEMat0_3}, where at each iteration $\beta$ is set according to the least squares approach, i.e.:
\begin{equation}
\label{eq:Transf4}
\begin{aligned}
\beta = \frac{\text{tr}\left\{\boldsymbol{\Theta} \mathbf{T}^H(\boldsymbol{\eta})\mathbf{A}^H\right\}}{\text{tr}\left\{\mathbf{A} \mathbf{T}(\boldsymbol{\eta}) \left(\mathbf{A} \mathbf{T}(\boldsymbol{\eta})\right)^H\right\}}.
\end{aligned}
\end{equation} 

\subsection{Results and comparisons}
The performance and convergence behavior of the gradient descent algorithm (GDA) described in \eqref{eq:mMSEMat0_3} depend on the initial conditions and the strategy for selecting the learning rate $\alpha$. Regarding the first issue, this paper considers random initial conditions, i.e., random initial values of $\boldsymbol{\eta}$, and for each case studied, we consider a multi-start approach with different initial conditions.  Hence, for the setting of $\alpha$, we adopted the backtracking line search strategy proposed in \cite{Boydbook}, which ensures convergence to a local minimum.

In the following results, we report the value of the normalized mean squared error $\epsilon ={\sum\limits_i \epsilon_i(\boldsymbol{\eta})}/{M^2}$, which represents the percentage error for each element of the 2D DFT transform. The results are obtained by running the GDA algorithm for a maximum of $10^5$ iterations, stopping the algorithm if $\epsilon \le 10^{-4}$.

Two different dimensions of the 2D DFT were considered: $DFT_1$ where $L_y = 4$ and $L_z = 2$, and $DFT_2$ where $L_y = 8$ and $L_z = 2$. Additionally, two different dimensions of the T-RIS constituting the SIM were examined: $T-RIS_1$ where $N_y = 16$ and $N_z = 4$, and $T-RIS_2$ where $N_y = 32$ and $N_z = 4$. 
Then, the cases $D-SIM$, which adheres to the general diagonal SIM model, and $DU-SIM_{id}$ with ideal T-RIS and unilateral approximation, were analyzed. Finally, the case where there is a mismatch between the model considered for optimization, assumed as $DU-SIM_{id}$, and the actual model used to calculate the results, assumed as $D-SIM$, was also considered and denoted $MDU-SIM_{id}$. To clarify further, the two cases $DU-SIM_{id}$ and $MDU-SIM_{id}$ utilize the same standard model to characterize the SIM and employ the same optimization algorithm, but they differ in their approach to testing results. In the $DU-SIM_{id}$ case, results are assessed with the assumption that the model remains identical to the one used for optimization. In contrast, the $MDU-SIM_{id}$ case leverages the complete model $D-SIM$ that does not incorporate the intrinsic approximations of the former model. This case allows for the verification of the applicability of the assumptions commonly made in the considered case study. A table summarizing the parameters of interest in the simulations and the case studies is presented in Table \ref{TableI}.}

\begin{figure}
    \centering
    \includegraphics[width=0.9\columnwidth]{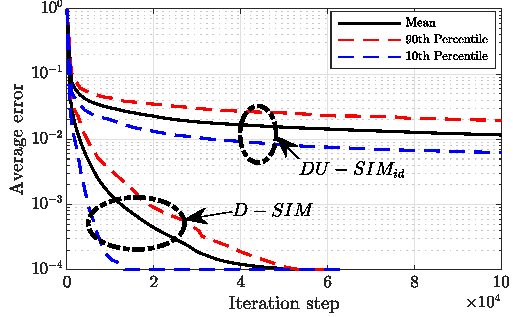} 
     \caption{$\epsilon$ as a function of the iterations of $DFT_1$ ($L_y = 4$, $L_z = 2$) with $T-RIS_1$ ($N_y = 16$, $N_z = 4$), $d_y = \lambda/2$, $d_x = \lambda$ and $Q = 3$.}
    \label{Fig1_new}
\end{figure}

\begin{figure}
    \centering 
    \includegraphics[width=0.9\columnwidth]
    {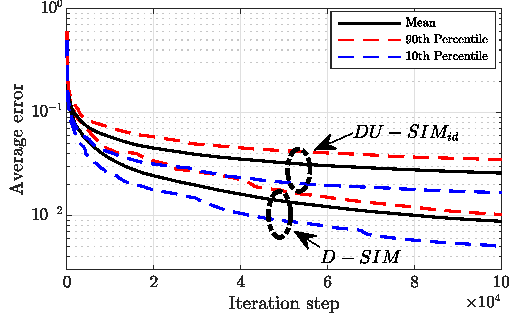}
   \caption{$\epsilon$ as a function of the iterations of $DFT_2$ ($L_y = 8$, $L_z = 2$) with $T-RIS_2$ ($N_y = 32$, $N_z = 4$), $d_y = \lambda/2$, $d_x = \lambda$ and $Q = 5$.}
    \label{Fig2_new}
\end{figure}

\begin{table}[h]
    \centering
    \small 
    \begin{tabular}{|c|c|c|}
        \hline
        \textbf{Parameter} & \textbf{Meaning} & \textbf{Value} \\ 
        \hline
        $f_0$ & Carrier frequency & 28 GHz \\ 
        \hline
        $L$ & Dipole length & $0.46 \lambda$ \\ 
        \hline
        $L_y$ & Probes in $y$ dir. & DFT\textsubscript{1}: 4 \\ 
        & & DFT\textsubscript{2}: 8 \\ 
        \hline
        $L_z$ & Probes in $z$ dir. & DFT\textsubscript{1}: 2 \\ 
        & & DFT\textsubscript{2}: 2 \\ 
        \hline
        $N_y$ & T-RIS dipoles in $y$ & T-RIS\textsubscript{1}: 16 \\ 
        & & T-RIS\textsubscript{2}: 32 \\ 
        \hline
        $N_z$ & T-RIS dipoles in $z$ & 4 \\ 
        \hline
        $d_x$ & Spacing between layers & $d_x = \lambda/2, \ \lambda, \ 2\lambda$ \\ 
        \hline
        $d_y$ & Spacing in $y$ dir. & $d_y = \lambda/3, \ \lambda/2, \ 2\lambda/3$ \\ 
        \hline
        $Q$ & Number of layers & $Q = 2, \ldots, 7$ \\ 
        \hline
    \end{tabular}
    \caption{Summary of parameters used in simulations.}
    \label{TableI}
\end{table}
\REV{In the first two figures, we illustrate the trend of $\epsilon$ as a function of the iterations for two specific cases, considering $d_y = \lambda/2$ and $d_x = \lambda$, which, as will be shown later, offer the best performance within a set of possible choices. The two figures present the curves of the average value of $\epsilon$ obtained from 100 instances of random generation of the initial conditions, along with the curves for the $10$th and $90$th percentiles for the cases $DU-SIM_{id}$ and $D-SIM$. Specifically, in Fig. \ref{Fig1_new}, we consider the case $DFT_1$ with $T-RIS_1$, $Q = 3$, and in Fig. \ref{Fig2_new}, we consider the case $DFT_2$ with $T-RIS_2$, $Q = 5$. 
It can be seen from the figures that if the complete SIM model without approximations is considered, namely in the case of $D-SIM$, better performance is achieved compared to when the simplified model $DU-SIM_{id}$ is used. This result highlights that the use of a complete multi-port network model, which takes into account the coupling among the elements and does not consider the unrealistic assumption of unidirectionality among the layers, not only leads to greater accuracy but also enables better performance. In the last figures, we will provide a graphical representation of the performance related to a single instance of the results from Fig. \ref{Fig2_new}, which will allow us to visualize the superiority of $D-SIM$ with respect to $DU-SIM_{id}$. The reason for this better behavior of $D-SIM$ may depend on the presence of couplings among the elements of the T-RIS and a feedback effect between the layers (due to not assuming unidirectionality), which allows for greater design flexibility. This result is valid also for the other cases considered which will be the subject of the upcoming figures. However, it should be noted that this result has been observed in the specific application scenario considered here, namely the implementation of a 2D DFT in the analog domain, and therefore we cannot claim it to hold in general. In any case, the main purpose of the results presented in this work is to demonstrate that SIM can provide the good results observed so far in various application scenarios, even when considering more accurate models to describe their functionalities. The results shown and those that follow indicate that for the test application scenario considered the goal has been achieved.}

\REV{In the next figure, we illustrate the behavior of the mean square error $\epsilon$, obtained at convergence or after a maximum of $10^5$ iterations, as a function of the number of levels $Q$ of the SIM for $d_y = \lambda/2$ and $d_x = \lambda$ for the case $DFT_2$ with $T-RIS_2$} 
The figure displays all the curves, namely $D-SIM$, $DU-SIM_{id}$, and $MDU-SIM_{id}$. 
The most important observations that can be drawn from the analysis of the figure is, on one hand, that performance improves with an increasing number of levels of the SIM. 
On the other hand, it is confirmed that the $D-SIM$ case shows a certain improvement in performance compared to the $DU-SIM_{id}$ case. \REV{Finally, we observe that, as expected, the case $MDU-SIM_{id}$ presents very poor results. This outcome is due to the mismatch that the assumptions typically made in the literature to characterize the SIM introduce in relation to the complete model, rendering these approximations inapplicable for optimizing the SIM. This result has been consistently observed across all the simulation setups considered in this study.}

\begin{figure}
    \centering
    \includegraphics{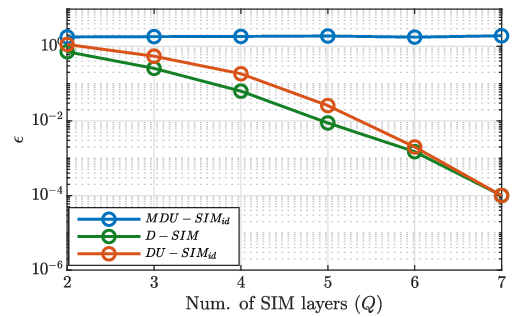}
    \caption{$\epsilon$ of $DFT_2$ ($L_y = 8$, $L_z = 2$) with $T-RIS_2$ ($N_y = 32$, $N_z = 4$).}
    \label{Fig:8probes}
\end{figure}
\begin{figure}
    \centering 
    \includegraphics[width=1\columnwidth]
    {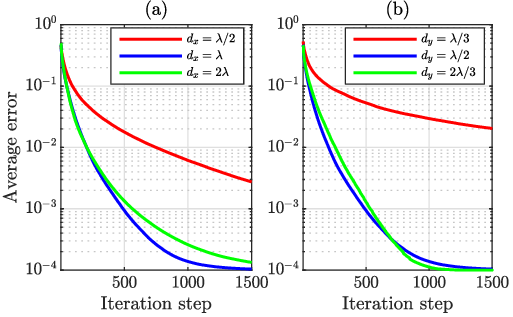}
   \caption{\REV{$\epsilon$ as a  function of the iterations for the $DFT_1$ case with the $D-SIM$ model, $Q = 5$ and (a): $d_y = \lambda/2$ and $d_x = \lambda/2$, $\lambda$, $2\lambda$; (b): $d_x = \lambda$ and $d_y = \lambda/3$, $\lambda/2$, $2\lambda/3$.}}
    \label{Fig8_new}
\end{figure}

In the next two figures, we show the trend of the value of $\epsilon$ as a function of the iterations, averaged over 100 iterations with random initial conditions for different values of $d_y$ and $d_x$ for the case $DFT_1$ with the $D-SIM$ model and $Q = 5$. Specifically, in Figure \ref{Fig8_new} (a), we present the results for $d_y = \lambda/2$ and $d_x = \lambda/2$, $\lambda$, $2\lambda$, while in Figure \ref{Fig8_new} (b), we present the results for $d_x = \lambda$ and $d_y = \lambda/3$, $\lambda/2$, $2\lambda/3$. The results highlight that the choice made in the results shown in the previous figures to consider $d_x = \lambda$ is the best among those considered, as it allows for faster convergence. Moreover, considering different $d_y$, it is observed that the cases $d_y = \lambda/2$ and $d_y = 2\lambda/3$ provide similar performance, while the case $d_y = \lambda/3$ performs significantly worse. Therefore, the choice $d_y = \lambda/2$ seems to be the most reasonable, as it allows for a reduction in footprint compared to the case $d_y = 2\lambda/3$. The results obtained for other parameters setting and for the $DFT_2$ case, substantially confirm this trend.

This behavior may be due to the fact that a too tight coupling between the elements of the SIM, which occurs by decreasing $d_y$, introduces excessive feedback that makes the convergence of the algorithm difficult or leads to poor local optima. Conversely, a loose coupling between the SIM layers, achieved by increasing $d_x$, makes the structure inherently less capable of realizing the desired transfer function. However, these considerations are valid for the specific case under study, namely the 2D DFT implementation, and cannot be generalized. They highlight that the optimization of the SIM architecture is an important topic that significantly impacts its performance and deserves further exploration, which will be the focus of future studies.

\begin{figure}
    \centering
    \includegraphics[width=0.54\textwidth]{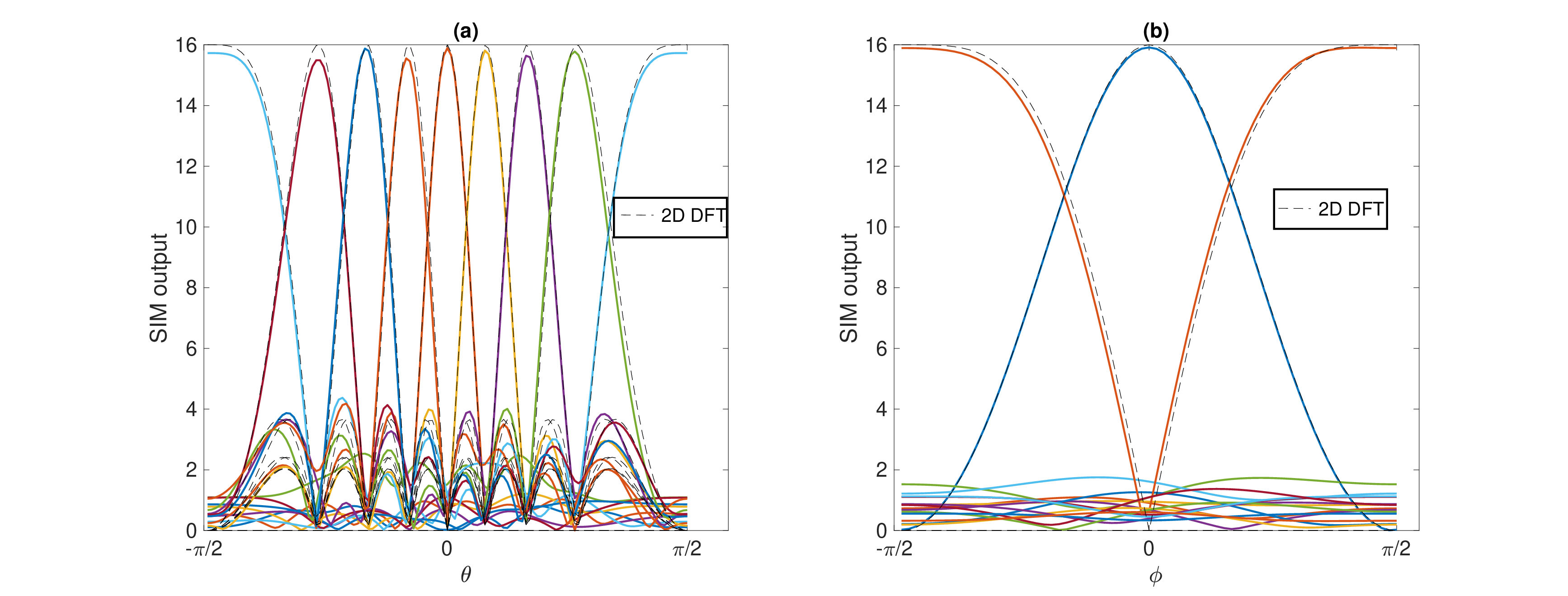}
    \caption{Response of all probes: $D-SIM$ with $Q = 4$.}
    \label{Fig:Probe_D_SIM_4}
\end{figure}

\begin{figure}
    \centering
    \includegraphics[width=0.54\textwidth]{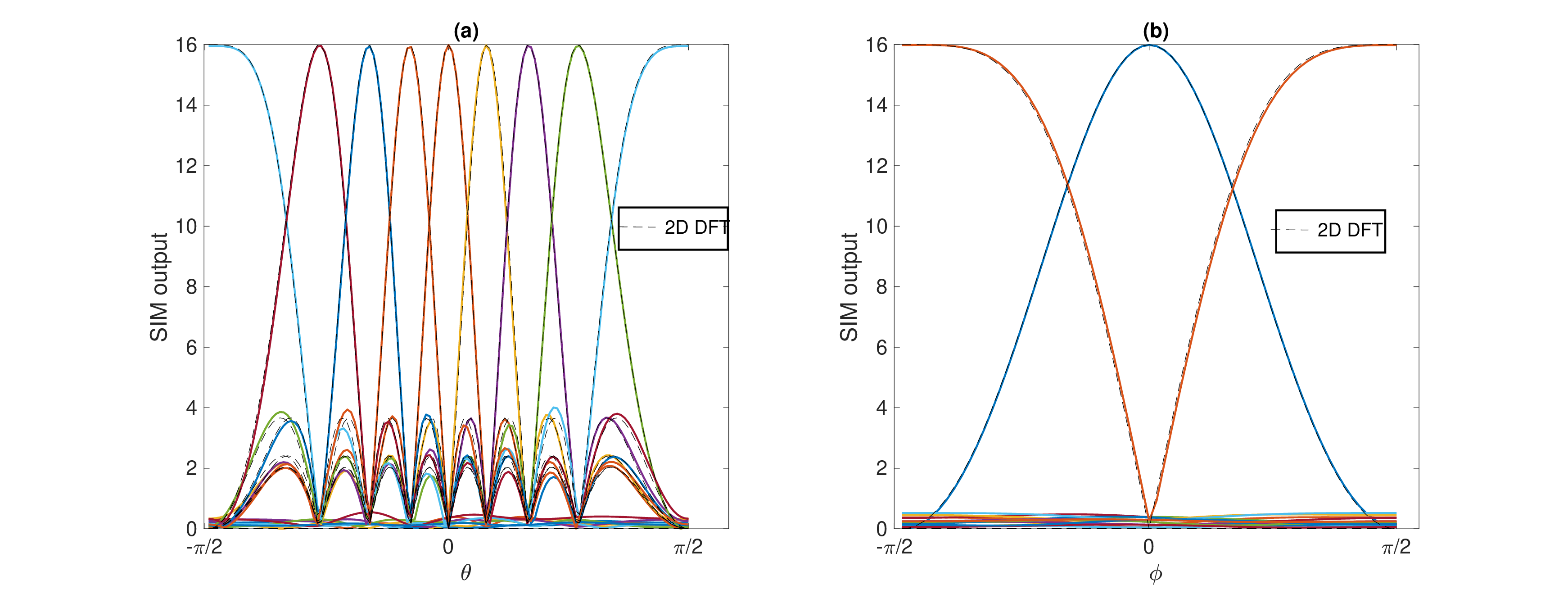}
    \caption{Response of all probes: $D-SIM$ with $Q = 5$.}
    \label{Fig:Probe_D_SIM_5}
\end{figure}

\begin{figure}
    \centering
    \includegraphics[width=0.54\textwidth]{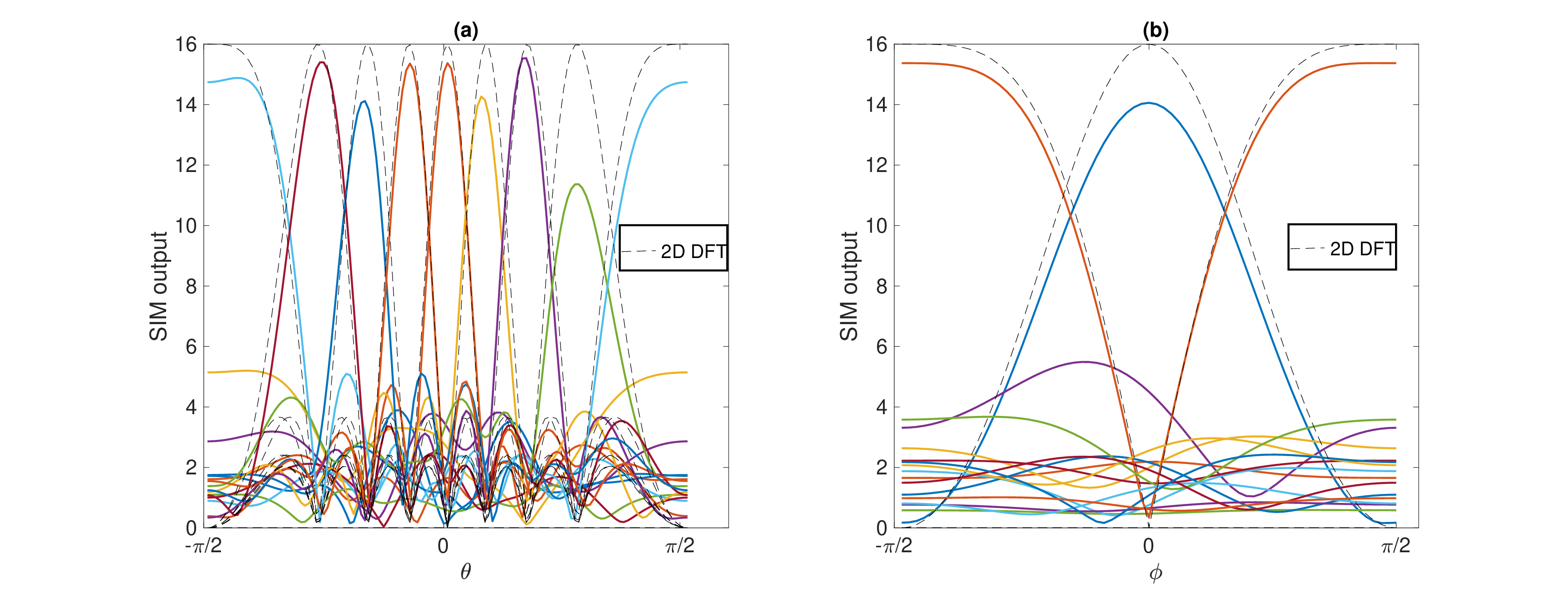}
    \caption{Response of all probes: $DU-SIM_{id}$ with $Q = 4$.}
    \label{Fig:Probe_DU_SIM_4}
\end{figure}

\begin{figure}
    \centering
    \includegraphics[width=0.54\textwidth]{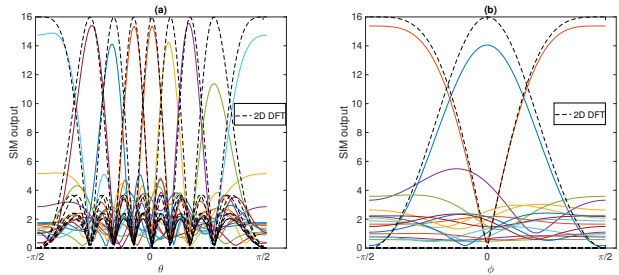}
    \caption{Response of all probes: $DU-SIM_{id}$ with $Q = 5$.}
    \label{Fig:Probe_DU_SIM_5}
\end{figure}

In order to provide a qualitative measure of the quality of the 2D DFT approximation, the following figures present a graphical representation of the SIM response to a plane wave \REV{for a single instance, specifically a random generation of the initial conditions of the algorithm}. Specifically, let $b_m$ denote the normalized input signal due to an incident plane wave coming from the azimuthal direction $\theta$ and with elevation $\phi$. We The input signal is:
\begin{equation}
\label{eq:input_plane_wave}
\begin{aligned}
{b}_m = e^{-j\frac{2\pi d_y}{\lambda}(m_y-1)\sin\theta}e^{-j\frac{2\pi d_z}{\lambda}(m_z-1)\sin\phi},
\end{aligned}
\end{equation}
for $m_y = 1,\ldots,L_y$ and $m_z = 1,\ldots,L_z$ and $m = (m_z-1)L_z + m_y$. As is well known, in this case, the 2D DFT is able to provide an estimate of the pair of angles of arrival $\theta$, $\phi$ simply by considering the pair of spatial frequencies for which the maximum magnitude of the 2D DFT is obtained. Then, we evaluate the magnitudes of the responses of all the probes for different angles $\phi \in [-\pi/2,\pi/2]$ and $\theta \in [-\pi/2,\pi/2]$, and the results are shown in figures \ref{Fig:Probe_D_SIM_4}, \ref{Fig:Probe_D_SIM_5}, \ref{Fig:Probe_DU_SIM_4}, and \ref{Fig:Probe_DU_SIM_5}. In all plots, the case $DFT_2$ with $T-RIS_2$ is considered. More specifically, the $D-SIM$ case with $Q = 4$ and $Q = 5$ is analyzed in Figs \ref{Fig:Probe_D_SIM_4} and \ref{Fig:Probe_D_SIM_5}, respectively, while the case $DU-SIM_{id}$ with $Q = 4$ and $Q = 5$ is analyzed in Figs \ref{Fig:Probe_DU_SIM_4} and \ref{Fig:Probe_DU_SIM_5}, respectively. In each figure, part (a) shows the responses as a function  of $\theta$ for $\phi = 0$, while part (b) displays the responses as a function of $\phi$ for $\theta = 0$. The dashed curves in each figure represent the ideal response, which is that of a 2D DFT transformation. The responses of the different probes are represented by different colors and indicate that, in all cases, the SIM allows for the activation of a probe at a specific angle, functioning as an estimator of the angle of arrival. However, it is evident that the $D-SIM$ case shows better performance for the same value of $Q$, meaning it provides a better approximation of the 2D DFT. The results obtained, although derived from a specific case, lend substantial support to the expectation that SIMs can realize a broad spectrum of processing functionalities for both communications and sensing applications.

\section{Conclusion}
In this work, we introduced a comprehensive multiport network model for the optimization of SIMs. In particular, by situating our approach within the context of a general Electromagnetic Collaborative Object (ECO), we established a foundational Z-parameter model that allows for the investigation of SIM architectures without relying on limiting assumptions, such as unilateral approximations or the absence of mutual coupling. 
Hence, we emphasized the impact of commonly used assumptions on model performance and potential simplifications, illustrating how variations in these assumptions can affect the complexity of the problem. \REV{Then, we have shown that the comprehensive model considered in this work reduces to the model traditionally used in the literature when the assumption of unilateral propagation between the levels of the SIM is made, and mutual coupling between the SIM elements is neglected. To assess the impact of these assumptions, a case study focused on the realization of a 2D DFT was considered. In this setting, we have shown that the mismatch caused by these assumptions renders the traditional model inadequate for optimizing the SIM. Conversely, we show that employing the complete model proposed in this paper can yield very good performance.}

\REV{The results obtained are very encouraging in the perspective of considering SIMs as one of the enabling technologies for future communication scenarios. In fact, the general theme of Integrated Sensing and Communications (ISAC) and the use of artificial intelligence (AI) have garnered increasing attention recently, driven by the forthcoming deployment of sixth-generation (6G) and subsequent communication systems. In this context, SIMs promise to provide a significant advantage over alternative approaches due to their ability to execute signal processing entirely in the electromagnetic domain. This may allow for a notable reduction in the costs associated with radio frequency chains and digital-to-analog converters, as well as computing costs. The findings of this study pave the way for innovative applications of SIM technology across various communication and sensing domains, promising enhancements in both performance and cost-effectiveness.}
\section*{\REV{Appendix A}}
Let us denote $\mathbf{T}_r = \mathbf{T}_{2Q,r}(\boldsymbol{\eta})$, with $r = 1,\ldots,2Q$, so that:
\begin{equation}
\label{eq:R1it_0}
\begin{aligned}
\mathbf{R}_q(\boldsymbol{\eta}) = \left\{\mathbf{T}_{2q-1},\mathbf{T}_{2q}\right\}.
\end{aligned}
\end{equation}
Since $\mathbf{T}(\boldsymbol{\eta})(\mathbf{Z}_{EE} + \mathbf{Z}_{E}(\boldsymbol{\eta})) = \mathbf{I}_N$, we have:
\begin{equation}
\label{eq:R1it_1}
\begin{aligned}
\mathbf{T}_{1}\left(\mathbf{W}^{(0)}_{2,2}+\mathbf{X}^{(1)}_{1,1}\right)+\mathbf{T}_{2} \mathbf{X}^{(1)}_{2,1} = \mathbf{0},
\end{aligned}
\end{equation}
and for $q=1,Q-1$:
\begin{equation}
\label{eq:R1it_2}
\begin{aligned}
& \mathbf{T}_{2q-1}\mathbf{X}^{(q)}_{1,2} + \mathbf{T}_{2q}\left(\mathbf{W}^{(q)}_{1,1}+\mathbf{X}^{(q)}_{2,2}\right)+\mathbf{T}_{2q+1} \mathbf{W}^{(q)}_{2,1} = \mathbf{0} \\
& \mathbf{T}_{2q}\mathbf{W}^{(q)}_{1,2} + \mathbf{T}_{2q+1}\left(\mathbf{W}^{(q)}_{2,2}+\mathbf{X}^{(q+1)}_{1,1}\right)+\mathbf{T}_{2q+2} \mathbf{X}^{(q+1)}_{2,1} = \mathbf{0},
\end{aligned}
\end{equation}
and:
\begin{equation}
\label{eq:R1it_3}
\begin{aligned}
\mathbf{T}_{2Q-1}\mathbf{X}^{(Q)}_{1,2} + \mathbf{T}_{2Q}\left(\mathbf{W}^{(Q)}_{1,1}+\mathbf{X}^{(Q)}_{2,2}\right) = \mathbf{I}_K.
\end{aligned}
\end{equation}
The relationships shown above allow the development of an iterative strategy for the calculation of the $\mathbf{T}_r$. Specifically, we introduce the matrices $\mathbf{M}_r \in \mathbb{C}^{K \times K}$, with $r = 0,\ldots,2Q$, defined as follows:
\begin{equation}
\label{eq:R1it_4}
\begin{aligned}
\mathbf{M}_0 & = \mathbf{0} \\
\mathbf{M}_1 & = \mathbf{I}_K, \\
\end{aligned}
\end{equation}
and for $q=1,Q-1$:
\begin{equation}
\label{eq:R1it_5}
\begin{aligned}
\mathbf{M}_{2q}  & = -\left[\mathbf{M}_{2q-2}\mathbf{W}^{(q-1)}_{1,2}+\mathbf{M}_{2q-1}\left(\mathbf{W}^{(q-1)}_{2,2}+\mathbf{X}^{(q)}_{1,1}\right)\right] \\
& \times \left(\mathbf{X}^{(q)}_{2,1}\right)^{-1}\\
\mathbf{M}_{2q+1}  & = -\left[\mathbf{M}_{2q-1}\mathbf{X}^{(q)}_{1,2}+\mathbf{M}_{2q}\left(\mathbf{W}^{(q)}_{1,1}+\mathbf{X}^{(q)}_{2,2}\right)\right] \left(\mathbf{W}^{(q)}_{2,1}\right)^{-1},
\end{aligned}
\end{equation}
and:
\begin{equation}
\label{eq:R1it_6}
\begin{aligned}
\mathbf{M}_{2Q}  & = -\left[\mathbf{M}_{2Q-2}\mathbf{W}^{(Q-1)}_{1,2}+\mathbf{M}_{2Q-1}\left(\mathbf{W}^{(Q-1)}_{2,2}+\mathbf{X}^{(Q)}_{1,1}\right)\right] \\
& \times \left(\mathbf{X}^{(Q)}_{2,1}\right)^{-1}.
\end{aligned}
\end{equation}
From the relationships \eqref{eq:R1it_1}, \eqref{eq:R1it_2}, and \eqref{eq:R1it_3}, it is therefore easy to derive:
\begin{equation}
\label{eq:R1it_7}
\begin{aligned}
\mathbf{T}_{1} & = \left[\mathbf{M}_{2Q-1}\mathbf{X}^{(Q)}_{1,2}+\mathbf{M}_{2Q}\left(\mathbf{W}^{(Q)}_{1,1}+\mathbf{X}^{(Q)}_{2,2}\right)\right]^{-1} \\
\mathbf{T}_{r} & = \mathbf{T}_{r-1} \mathbf{M}_{r} \quad \text{for $r = 2,\ldots,2Q$}.
\end{aligned}
\end{equation}
As for the evaluation of $\mathbf{S}_q(\boldsymbol{\eta})$, let us denote $\mathbf{U}_r = \boldsymbol{T}_{r,1}(\boldsymbol{\eta})$, with $r = 1,\ldots,2Q$, so that:
\begin{equation}
\label{eq:C1it_0}
\begin{aligned}
\mathbf{S}_q(\boldsymbol{\eta}) = \left\{\mathbf{U}^T_{2q-1},\mathbf{U}_{2q}^T\right\}.
\end{aligned}
\end{equation}
From the definition of $\mathbf{U}_r$ and $\mathbf{T}_r$ we have:
\begin{equation}
\label{eq:C1it_1}
\begin{aligned}
\mathbf{U}_{2Q} = \mathbf{T}_{1}.
\end{aligned}
\end{equation}
Since $(\mathbf{Z}_{EE} + \mathbf{Z}_{E}(\boldsymbol{\eta}))\mathbf{T}(\boldsymbol{\eta}) = \mathbf{I}_N$, it is easy to derive the following iterative procedure:
\begin{equation}
\label{eq:C1it_2}
\begin{aligned}
\mathbf{U}_{2Q-1} = -\left(\mathbf{X}^{(Q)}_{2,1}\right)^{-1} \left(\mathbf{W}^{(Q)}_{1,1}+\mathbf{X}^{(Q)}_{2,2}\right)\mathbf{U}_{2Q},
\end{aligned}
\end{equation}
and for $q = 1,\ldots,Q-1$:
\begin{equation}
\label{eq:C1it_3}
\begin{aligned}
& \mathbf{U}_{2Q-2q} = -\left(\mathbf{W}^{(Q-q)}_{2,1}\right)^{-1} \\
& \times \left[\left(\mathbf{W}^{(Q-q)}_{2,2}+\mathbf{X}^{(Q-q+1)}_{1,1}\right)\mathbf{U}_{2Q-2q+1}+\mathbf{X}^{(Q-q+1)}_{1,2}\mathbf{U}_{2Q-2q+2}\right]\\
& \mathbf{U}_{2Q-2q-1} = -\left(\mathbf{X}^{(Q-q)}_{2,1}\right)^{-1} \\
& \times \left[\left(\mathbf{W}^{(Q-q)}_{1,1}+\mathbf{X}^{(Q-q)}_{2,2}\right)\mathbf{U}_{2Q-2q}+\mathbf{W}^{(Q-q)}_{1,2}\mathbf{U}_{2Q-2q+1}\right].
\end{aligned}
\end{equation}
\section*{\REV{Appendix B}}
Given the assumption $\mathbf{W}^{(q)}_{1,2} = \mathbf{0}$, it is possible to write the pair of relationships:
\begin{equation}
\label{eq:Uni1}
\begin{aligned}
\mathbf{T}_{2Q-1}\mathbf{X}^{(Q)}_{1,2}  +\mathbf{T}_{2Q}\left(\mathbf{X}^{(Q)}_{2,2} +\mathbf{W}^{(Q)}_{1,1}\right) & = \mathbf{I}_K\\
\mathbf{T}_{2Q-1}\left(\mathbf{X}^{(Q)}_{1,1}+\mathbf{W}^{(Q-1)}_{2,2}\right)  +\mathbf{T}_{2Q}\mathbf{X}^{(Q)}_{2,1} & = \mathbf{0}.
\end{aligned}
\end{equation}
Introducing 
\begin{equation}
\label{eq:Uni5}
\begin{aligned}
\boldsymbol{\Omega}_{p} = \left(\mathbf{X}^{(q)}_{2,2} +\mathbf{W}^{(p)}_{1,1}-\mathbf{X}^{(p)}_{2,1}\left(\mathbf{X}^{(p)}_{1,1}+\mathbf{W}^{(p-1)}_{2,2}\right)^{-1}\mathbf{X}^{(p)}_{1,2}\right)^{-1},
\end{aligned}
\end{equation}
from \eqref{eq:Uni1} we can directly evaluate $\mathbf{T}_{2Q}$ as:
\begin{equation}
\label{eq:Uni2}
\begin{aligned}
\mathbf{T}_{2Q} & = \boldsymbol{\Omega}_{Q}.
\end{aligned}
\end{equation}
Then, we have for $q = Q,Q-1,\ldots,2$: 
\begin{equation}
\label{eq:Uni3}
\begin{aligned}
\mathbf{T}_{2q-1}  & = - \mathbf{T}_{2q} \mathbf{X}^{(q)}_{2,1}\left(\mathbf{X}^{(q)}_{1,1}+\mathbf{W}^{(q-1)}_{2,2}\right)^{-1} \\ 
\mathbf{T}_{2q-2}  & = - \mathbf{T}_{2q-1} \mathbf{W}^{(q-1)}_{2,1}\boldsymbol{\Omega}_{q-1},
\end{aligned}
\end{equation}
and
\begin{equation}
\label{eq:Uni4}
\begin{aligned}
\mathbf{T}_{1}  & = - \mathbf{T}_{2} \mathbf{X}^{(1)}_{2,1}\left(\mathbf{X}^{(1)}_{1,1}+\mathbf{W}^{(0)}_{2,2}\right)^{-1}.
\end{aligned}
\end{equation}

Regarding the computation of $\mathbf{U}_{r}$, $r = 1,\ldots,2Q$, let us introduce the following:
\begin{equation}
\label{eq:Uni_c_5}
\begin{aligned}
\boldsymbol{\zeta}_{q} = \left(\mathbf{X}^{(q)}_{1,1} +\mathbf{W}^{(q-1)}_{2,2}-\mathbf{X}^{(q)}_{1,2}\left(\mathbf{X}^{(q)}_{2,2}+\mathbf{W}^{(q-1)}_{2,2}\right)^{-1}\mathbf{X}^{(q)}_{2,1}\right)^{-1}.
\end{aligned}
\end{equation}
It is easy to show in this case:
\begin{equation}
\label{eq:Uni_c_2}
\begin{aligned}
\mathbf{U}_{1} & = \boldsymbol{\zeta}_{1}, 
\end{aligned}
\end{equation}
and for $q = 1,2,\ldots,Q-1$: 
\begin{equation}
\label{eq:Uni_c_3}
\begin{aligned}
\mathbf{U}_{2q}  & =- \left(\mathbf{X}^{(q)}_{2,2}+\mathbf{W}^{(q)}_{1,1}\right)^{-1}\mathbf{X}^{(q)}_{2,1} \mathbf{U}_{2q-1} \\ 
\mathbf{U}_{2q+1}  & =- \boldsymbol{\zeta}_{q+1}\mathbf{W}^{(q)}_{2,1} \mathbf{U}_{2q}.
\end{aligned}
\end{equation}
\bibliographystyle{IEEEtran}


\end{document}